\documentclass{aa}  

\usepackage{graphicx}
\usepackage{txfonts}
\def\msun{M$_{\odot}$}
\def\Msun{M$_{\odot}$ }
\def\be{\begin{equation}}
\def\ee{\end{equation}}
\def\pd{\partial}
\def\ji{\varphi}
\begin{document}

   \title{The first direct double neutron star merger detection: implications for cosmic nucleosynthesis.}

\author{S. Rosswog\inst{1}
  \and J. Sollerman\inst{1} 
  \and U. Feindt\inst{2}
  \and A. Goobar\inst{2}
  \and O. Korobkin\inst{3}
  \and R. Wollaeger\inst{3}
  \and C. Fremling\inst{4}
  \and M. M. Kasliwal\inst{4}
} 

   \institute{ The Oskar Klein Centre, Department of Astronomy, AlbaNova, Stockholm University, SE-106 91 Stockholm, Sweden
         \and
             The Oskar Klein Centre, Department of Physics, AlbaNova, Stockholm University, SE-106 91 Stockholm, Sweden
         \and Los Alamos National Laboratory, Los Alamos, NM 87545, USA
         \and Division of Physics, Mathematics and Astronomy, California Institute of Technology, Pasadena, CA 91125, USA
             }

   \date{Received October, 2017; accepted }

 
  \abstract
   {The astrophysical r-process site where about half of the 
   elements heavier than iron are produced has been a puzzle for
    several decades. Here we discuss the role of one of the 
   leading ideas --neutron star mergers (NSMs)-- in the light of the 
   first direct detection of such an event in both gravitational 
   (GW) and electromagnetic (EM) waves.}
   {
    Understanding the implications of the first GW/EM observations 
    of a neutron star merger for cosmic nucleosynthesis.    
   }
   {We analyse bolometric and NIR lightcurves of the first 
   detected double neutron star merger and compare them to 
   nuclear reaction network-based macronova models.}
   {
   The slope of the bolometric lightcurve is consistent with the radioactive 
   decay of neutron star ejecta with $Y_e \lesssim 0.3$ (but not larger), 
   which provides strong evidence for an r-process origin of the electromagnetic 
   emission.  This rules out in particular "nickel winds" as major source of the 
   emission.
   We find that the NIR lightcurves can be well fitted either with or 
   without lanthanide-rich ejecta.
   Our limits on the ejecta mass together with estimated rates 
   directly confirm earlier purely theoretical or indirect observational 
   conclusions that double neutron star mergers are indeed a major site 
   of cosmic nucleosynthesis. If the ejecta mass was {\em typical}, NSMs
   can easily produce {\em all} of the estimated Galactic r-process matter,
   and --depending on the real rate-- potentially even more. This could be a 
   hint that the event ejected a particularly large amount of mass, maybe due to a 
   substantial difference between the component masses.
   This would be compatible with the mass limits obtained from the 
   GW-observation.
   }
   {The recent observations suggests that NSMs are responsible 
    for a broad range of r-process nuclei and that they are at least a major, but
     likely the dominant r-process site in the Universe.
}


   \maketitle

\section{Introduction}

Soon after the discovery 
of the first binary neutron star \citep[PSR 1913+16;][]{hulse75} 
it became clear that gravitational wave emission drives the  binary system towards a final
coalescence \citep{taylor82}. 
\cite{lattimer74} speculated that neutron star debris from such an encounter could be a promising production
site\footnote{For technical reasons they performed the analysis for a neutron star - black hole 
system.} for the heaviest elements formed via "rapid neutron capture" or "r-process" 
\citep{burbidge57,cameron57,thielemann11}. With the techniques available at that time
they could, however, only estimate the ejecta mass to  "$\sim$ 0.05 $\pm$ 0.05 M$_{\rm ns}$".
The  r-process is responsible for about half of the 
elements heavier than iron, but  until recently the dominant opinion was that core-collapse 
supernovae (CC SNe) must be the major production site. \cite{eichler89} discussed merging neutron star 
binaries as "central engines" for short gamma-ray bursts (sGRBs) and as r-process production 
sites. The first  nucleosynthesis calculations based on 3D hydrodynamic merger simulations \citep{rosswog99}
showed  that the neutron-rich matter that is dynamically ejected  indeed produces 
--robustly and without any fine-tuning-- r-process nuclei up to and beyond 
the third r-process peak at nucleon numbers of $A=195$ 
\citep{rosswog98b,freiburghaus99b}. They also showed that
the ejecta are --if folded with estimated merger rates-- enough 
to explain the amount of r-process material in the Galaxy.\\
A large number of subsequent studies \citep[e.g.][]{roberts11, goriely11a,wanajo12,korobkin12a}, 
have investigated these so-called "dynamic ejecta" 
as r-process sites. Only more recently, it was realised that the extremely
low electron fraction (= electron to nucleon ratio= proton to nucleon ratio) ejecta ($Y_e \lesssim 0.1$) 
are likely complemented by matter reaching 
$Y_e \sim 0.3$, e.g. by shock-heated material \citep{wanajo14,radice16a},
neutrino-driven winds \citep{dessart09,perego14b} or the unbinding of accretion torus material 
\citep[e.g.][]{lee07,beloborodov08,metzger08a,fernandez13a,just15,ciolfi15,martin15,siegel17a}. 
This unbound torus material can amount to $\sim40\%$ of the initial 
torus mass and --depending on the initial mass asymmetry-- 
can actually dominate the ejecta. Geometrically, there is the tendency of the low-$Y_e$ matter to
be concentrated towards the orbital plane, while $Y_e$ increases towards the polar remnant regions,
 see e.g. Figs. 14 and 15 in \cite{perego14b}.\\
While initially questioned \citep[e.g.][]{argast04},
a number of recent studies \citep{matteucci14a,mennekens14a,vandevoort14a,shen15a} find compact 
binary mergers at least as suitable or even preferred over CC SNe as the major r-process production 
site. One of the differences between the main alternatives is that 
(at least "ordinary") CC SNe occur $\sim 1000$ more frequently than compact binary mergers
and therefore have to deliver a correspondingly smaller amount of r-process elements per event to account for
the cosmic inventory. There are, however, various lines of arguments that favour rare events with large ejecta
masses over frequent occurrences with smaller ones. For example, the geochemical enrichment of 
$^{244}$Pu \citep{wallner15a,hotokezaka15a} and the observation of r-process enriched ultra-faint 
dwarf galaxies \citep{beniamini16a,hansen17a} both argue in favour of rare events with high mass ejection.
The inferred rates and ejecta masses agree well with what is expected from NSMs.\\
The most direct confirmation of compact binary mergers as r-process sites, however, is the
detection of electromagnetic radiation from the radioactive decay of freshly synthesised r-process
elements in the aftermath of a merger, a so-called "macronova" or "kilonova" 
\citep[e.g.][]{li98,kulkarni05,rosswog05a,metzger10b,kasen13a,tanaka13a}. The most compelling
previous evidence for such a macronova has been the detection of an infrared excess in the aftermath
of a short GRB (130603B; \citealt{tanvir13a,berger13b}).\\
The situation changed fundamentally on August 17, 2017 with the first direct detection of gravitational 
waves from a neutron star binary, GW170817 by the LIGO-Virgo collaboration \citep[LVC;][]{LIGO17_GW_NSNS}. 
The electromagnetic follow-up of GW170817 has been described in many papers
\citep[e.g.][]{ligo_NSNS_MultiMessenger17,kasliwal17,smartt17}, and includes the first detection in gamma-rays
\citep{fermiGBMpaper} only  1.7 seconds after GW170817, via the optical and near-infrared (NIR) discovery 
and monitoring of AT2017gfo \citep{ligo_NSNS_MultiMessenger17} to the late onset of the radio emission
\citep{vlapaper17}. The object was initially surprisingly bright and blue compared to pre-discovery 
predictions, it was discovered at an absolute magnitude of $-15.7$ in the i band
\citep{coulter17}. The source quickly declined in the optical bands and over the next 
3 weeks was observed to decline in the NIR, overall in agreement with the family of macronova 
models presented by e.g. \citet[][]{kasen13a,tanaka13a,rosswog17a,wollaeger17a}.
This detection marks the beginning of the long-awaited era in multi-messenger astronomy.\\
A large variety of facets and implications of this event have been discussed in the recent literature 
\citep[e.g.][]{abbott17_MM,abbott17b,evans17,kasen17a,levan17,tanvir17,margutti17}.
In this paper we focus on the implications of this first discovery for the cosmic nucleosynthesis. We 
will in particular analyze bolometric and near-infrared lightcurves with respect to what they imply 
for the ejecta parameters and their nucleosynthesis. We discuss the relevance of the inferred 
ejecta amount for the Galactic r-process inventory and update our recent predictions \citep{rosswog17a}
for the detectability of AT2017gf-like events.


\section{Bolometric lightcurve: a clue to r-process nucleosynthesis}
\label{sec:heating}
\begin{figure*}
\centerline{
\includegraphics[width=2.\columnwidth]{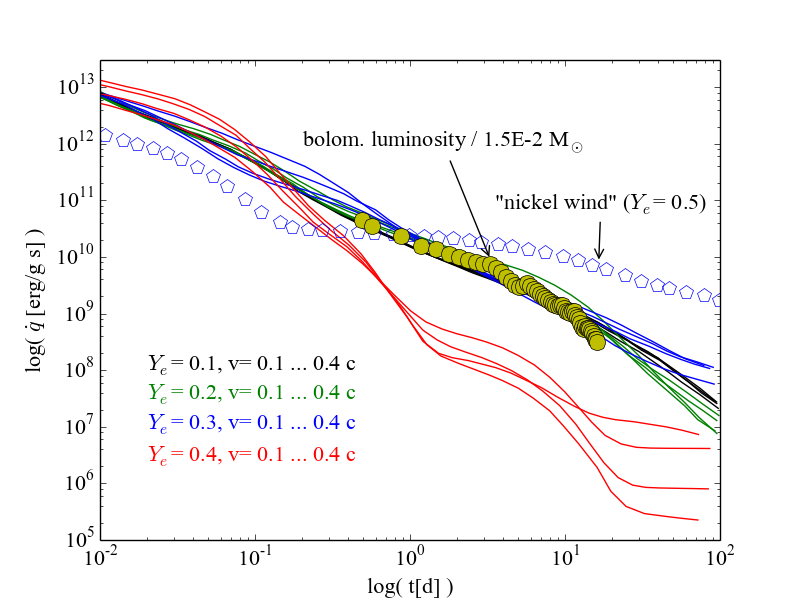}}
\vspace*{-0.2cm}
\caption{Nuclear heating rates of the explored parameter space, colours label $Y_e$-values. 
Overlaid are bolometric luminosities computed following the description in 
Kasliwal et al. (2017) using updated photometry from https://kilonova.space (yellow circles). 
We show the total nuclear heating rate (luminosities divided by an ejecta mass of 
 $1.5 \times 10^{-2}$ \msun). Also shown is the heating rate of a wind 
 with $Y_e= 0.5$ that produces a substantial amount of nickel, see last panel
 in Fig.~\ref{fig:nucleo_all_yes}.
 The close agreement with $Y_e \lesssim 0.3$ strongly  suggests the presence 
 of substantial amounts of r-process matter.}
\label{fig:heating}
\end{figure*}
\begin{figure*}
\vspace*{0cm}
\centerline{\includegraphics[width=2.\columnwidth,angle=0]{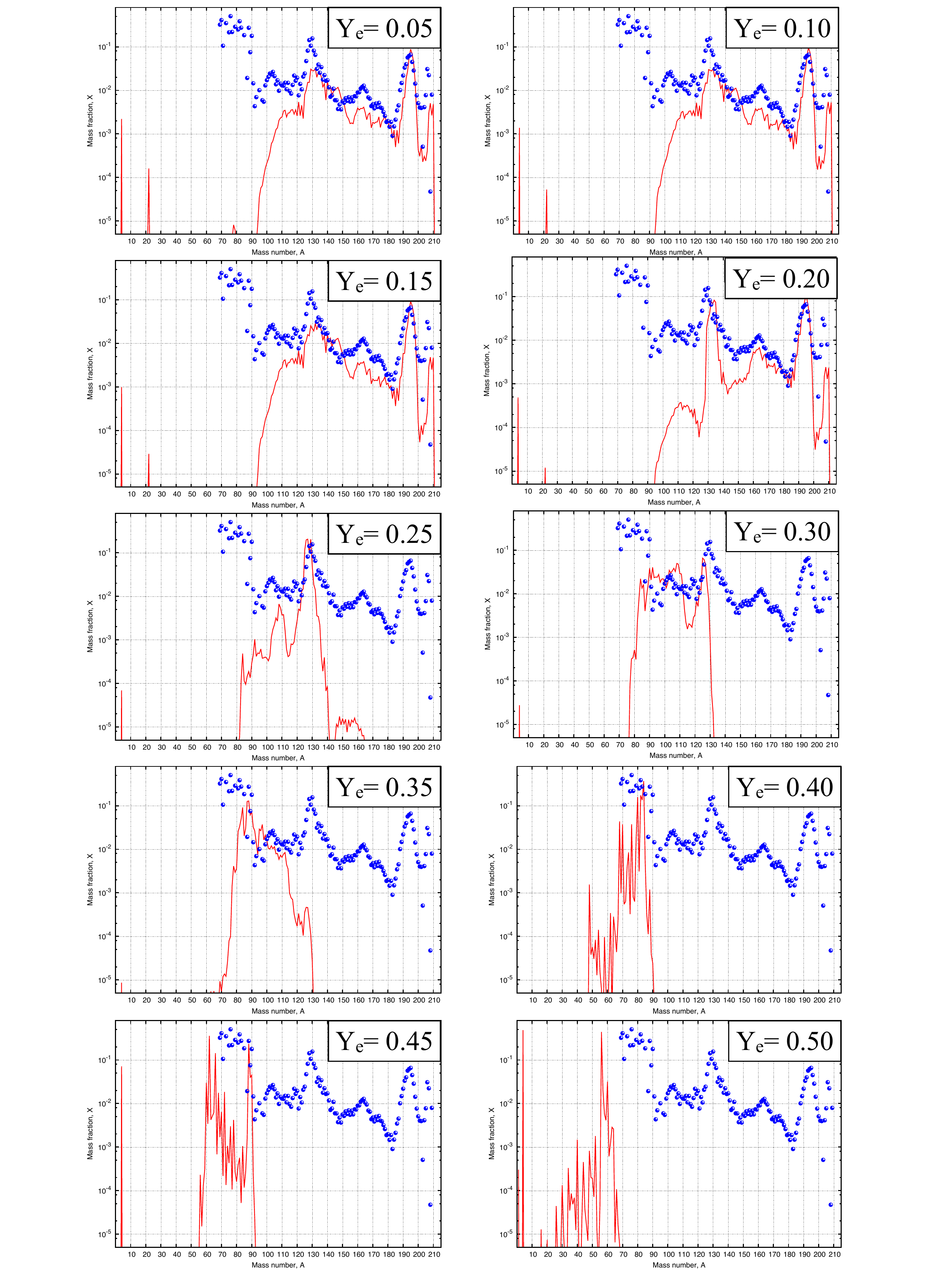}}
\vspace*{0cm}
\caption{Nucleosynthesis for different electron fractions $Y_e$, for $s_0= 15 k_B$/baryon, $v_{\rm ej}= 0.25$ c
and FRDM mass model.
Beyond $Y_e^{\rm crit} \approx 0.25$ hardly any heavy elements beyond the second r-process peak ($A= 130$)
are produced. Blue symbols refer to the solar system r-process.}
\label{fig:nucleo_all_yes}
\end{figure*}
\begin{figure}
\vspace*{-2.cm}
\centerline{\includegraphics[width=1\columnwidth,angle=0]{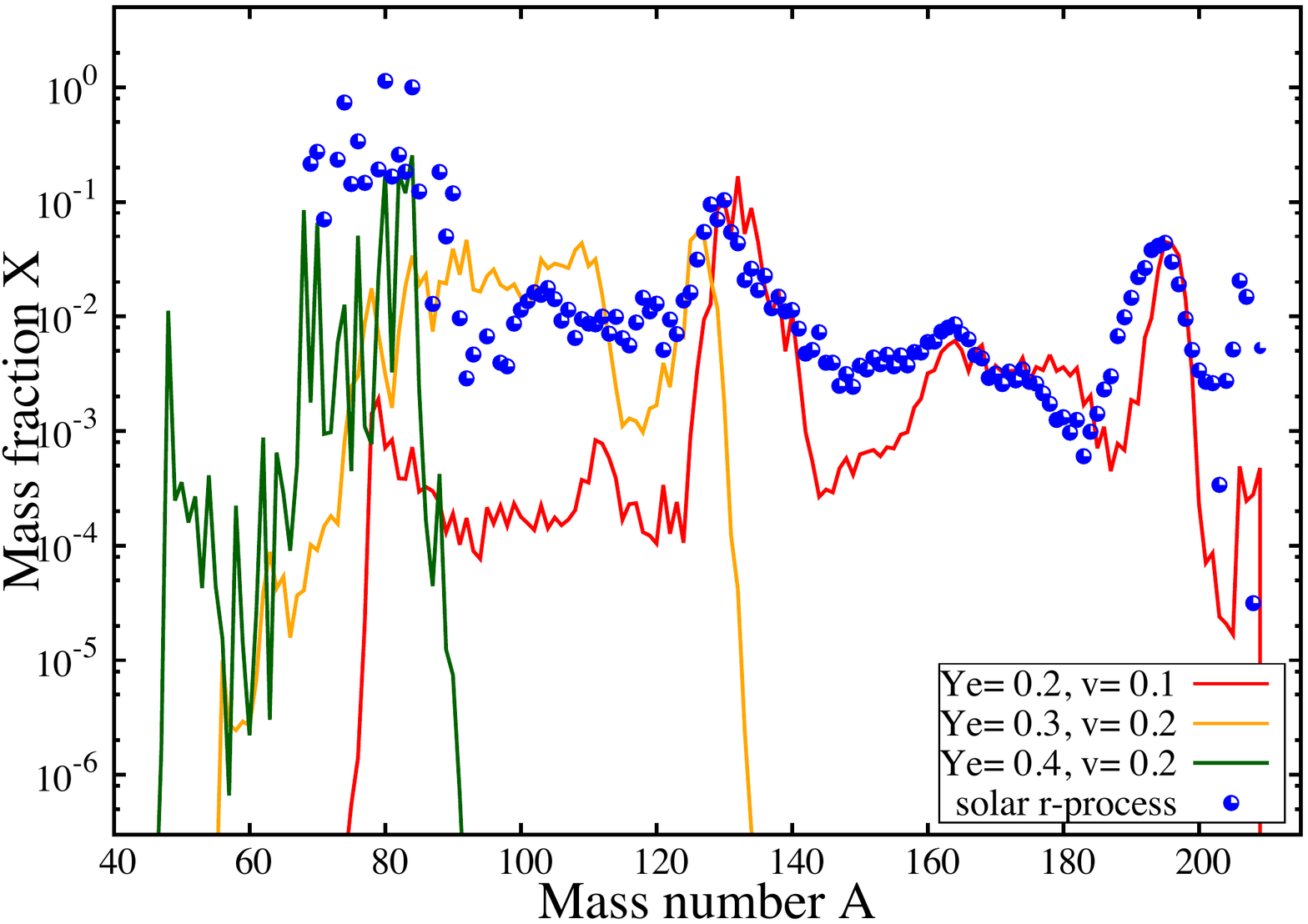}}
\vspace*{-2.5cm}
\caption{Abundances for two cases (red and orange lines) that can both reproduce the bolometric luminosity evolution. 
While both produce r-process material, one case produces the third r-process 
peak
($Y_e=0.2$ and $v=0.1$c),
but the other ($Y_e=0.3$ and $v=0.2$c) does not. 
Thus, from the bolometric lightcurve alone the
absence/presence of lanthanides cannot be inferred. For comparison, we  also show a case (green line) with
large $Y_e= 0.4$ ($v=0.2$c) that only produces elements with $A<90$.
That case does not fit the bolometric luminosity.
}
\label{fig:nucleo_best_bolometric}
\end{figure}

We have performed a number of calculations with the nuclear reaction network
WinNet \citep{winteler12,winteler12b} to explore how sensitive the nuclear heating rates are
to the physical expansion conditions, which we set up as described in 
\citet[][ their Sect. 2.2]{rosswog17a}. 
We ran a grid of 16 expansion models covering a broad parameter range 
($[v/c] \times [Y_e] = [0.1, 0.2, 0.3, 0.4] \times [0.1, 0.2, 0.3, 0.4]$). To keep the parameter space manageable we fix the initial 
entropy to 15 $k_B$. This is reasonable 
since a) for very low $Y_e$-values the results 
are insensitive to the exact entropy-value \citep{freiburghaus99a} and b)  for higher $Y_e$ 
cases detailed simulation studies find narrow distributions around this value 
\citep{perego14b,radice16a}. For each case a power-law approximation for the
nuclear heating rate (in erg/g s) 
\be
 \dot{q}= \dot{q}_0 \left( \frac{t}{t_0} \right)^\alpha
\ee
was determined from the network data (for $t> 10^{-4}$ d; at earlier times the heating
rate is roughly constant), see Table~\ref{tab:heating}.
We find that the power-law index $\alpha \approx -1.3$ for as long as $Y_e \lesssim 0.3$, 
consistent with earlier findings \citep{metzger10b,korobkin12a,hotokezaka17a}
When $Y_e= 0.4$ the heating rate drops off substantially faster, 
and the normalisation constant $\dot{q}_0$ is typically an order of magnitude 
lower\footnote{Obviously, for this case a single power law is not a good approximation.}. 
At early times when opacity effects are significant, diffusion can
substantially affect the lightcurve shape. Once the ejecta are optically thin, 
and excess radiation produced earlier had time to escape,
the lightcurve slope is determined by the heating rate from radioactive decay, 
modulo heating efficiency, $f_{\rm tot}$. 
We compare the heating rate with the bolometric luminosity where we use
data from  the Kilonova catalog\footnote{
https://kilonova.space/ retreived on Jan. 18, see \cite{guillochon17}
}. The bulk of the NIR data come from
\cite{smartt17}, \cite{kasliwal17} and \cite{tanvir17} as discussed in
\cite{villar17}, and additional g-band data mainly come from
\cite{pian17,arcavi17,coulter17} and \cite{cowperthwaite17}.
We find that the slope of the bolometric luminosity (yellow filled circles) agrees excellently with 
the one of the nuclear heating rates $\dot{q}$, provided that the electron fraction 
$Y_e\lesssim 0.3$, but not larger. Along such low-$Y_e$ trajectories 
r-process elements are forged, see Fig.~\ref{fig:nucleo_all_yes}, and therefore
the excellent agreement with the observed bolometric luminosity
strongly suggests an r-process origin of the observed emission. 
As a further comparison, we also plot a trajectory
with $Y_e= 0.5$ which produces a substantial amount of nickel (marked
with open polygons; same trajectory that produces the abundance pattern shown
in the last panel of Fig.~\ref{fig:nucleo_all_yes}). This rules out a
nickel wind as the primary source of bolometric luminosity. \\
To illustrate the impact of the electron fraction $Y_e$ on the resulting
abundance pattern, we plot in Fig.~\ref{fig:nucleo_all_yes}
the results for typical ejecta conditions ($s=15$ k$_{\rm B}$, $v_{\rm ej}= 0.25$ c)
where we systematically vary $Y_e$ from 0.05 to 0.5. Below
$Y_e^{\rm crit}\approx 0.25$ "heavy" r-process including lanthanides up to and
beyond the "platinum peak" at $A=195$ are produced, with a very robust abundance
pattern for $Y_e \lesssim 0.15$. Above $Y_e^{\rm crit}$ r-process still occurs, but
produces only light r-process elements ($A \lesssim 130$).
 At the high $Y_e$-end substantial amounts of nickel
are produced.
\\
The efficiency with which
released energy is translated into electromagnetic emission
depends on the detailed decay products \citep{barnes16a}, and varies
from $\sim 0.7$ at 1 day to $\sim 0.3$ at 20 days, see Fig.~8 in \cite{rosswog17a}.
Therefore, the {\em net} heating rates ($\dot{q} f_{\rm tot}$) decay slightly faster than the "naked" 
ones,  though still in very good agreement with the observed 
bolometric lightcurve.\\
Assuming that 100\% of the radioactive 
energy ends up in the observed emission places a lower limit
on the ejected mass of 
\be
m_{\rm ej}^{\rm min}\equiv \frac{L_{\rm bol}}{\dot{q}}
\approx 1.5 \times 10^{-2} {\rm M}_\odot.
\label{eq:ejected_mass}
\ee 
For a fixed set of nuclear physics ingredients this lower limit is
robust. It has, however, been stressed by both
\cite{barnes16a} and \cite{rosswog17a} that different 
nuclear mass models yield different amounts of trans-lead nuclei,
the decays of which can substantially enhance the nuclear heating rate. 
For example, the results for the Finite-Range Droplet Model 
\citep[FRDM; this mass model is used in WinNet;][]{moeller95} and 
the nuclear mass model of Duflo and Zuker \citep[DZ;][]{duflo95}
differed at time scales of about a day by a factor of $\sim 5$ in their 
net heating rates $\dot{q} f_{\rm tot}$. Therefore, {\em if} a large fraction of the ejecta would 
have an electron fraction $<0.25$ {\em and} the nuclear heating would
be close to the DZ-predictions, this mass limit could be smaller by
a factor of $\sim5$.\\
Since the bolometric light curve seems equally well fit by all the models 
with electron fractions $Y_e \lesssim 0.3$, but only material with 
$Y_e < Y_e^{\rm crit} \approx 0.25$  produces
the third r-process peak, see Fig.~(\ref{fig:nucleo_all_yes}), the bolometric luminosities
alone are not conclusive regarding the ejecta composition.
In particular it does not allow to infer whether lanthanides are 
present or whether the third r-process peak with elements such as platinum
or gold is produced.
For the purpose of illustration, we plot in Fig.~\ref{fig:nucleo_best_bolometric} 
the resulting abundances for three trajectories. The first two yield an 
excellent fit to the slope of the bolometric light curve, but one 
($Y_e=0.2$, $v= 0.1c$) produces the full r-process range (but abundances 
below the second peak are produced only sub-dominantly) while the other
($Y_e=0.3$, $v= 0.2c$) does not produce r-process beyond nucleon numbers 
$A>130$. 
For comparison we also show the abundances for $Y_e=0.4$ case which 
produces only elements up to $A\approx 90$.
\begin{table}
\begin{tabular}{@{}ccccccl@{}}
\hline
   $v [c]$    &  $Y_e$ & $\dot{q}_0$ [$10^{10}$erg(~g~s)$^{-1}$]                    & $\alpha$\\
   \hline \\
0.1  & 0.1 & $1.74 $ & -1.31\\
       & 0.2 & $2.14 $ & -1.28\\
       & 0.3 & $2.23 $ & -1.31\\
       & 0.4 & $0.234  $   & -1.66\\
  \\
0.2  & 0.1 & $1.80 $ &-1.31\\
       & 0.2 & $1.75 $ &-1.31\\
       & 0.3 & $2.48 $ &-1.27\\
       & 0.4 & $0.140 $   &-1.77\\
  \\
0.3  & 0.1 & $1.88    $ &-1.30\\
       & 0.2 & $1.67  $ &-1.31\\
       & 0.3 & $2.35  $ &-1.25\\
       & 0.4 & $0.104 $   &-1.81\\
  \\
0.4  & 0.1 & $1.92    $  &-1.30\\
       & 0.2 & $1.73  $ &-1.30\\
       & 0.3 & $2.28  $ &-1.23\\
       & 0.4 & $0.289 $   &-1.64\\
\end{tabular}
\caption{Coefficients for power-law fits for nuclear heating 
rates of the form $ \dot{q}= \dot{q}_0 \left( \frac{t}{t_0} \right)^\alpha$,
where $t_0$= 1 day.}
\label{tab:heating}
\end{table}

\section{Late near-infrared lightcurves}
\label{sec:NIR}
\begin{figure*}
\includegraphics[width=1.\columnwidth,angle=0]{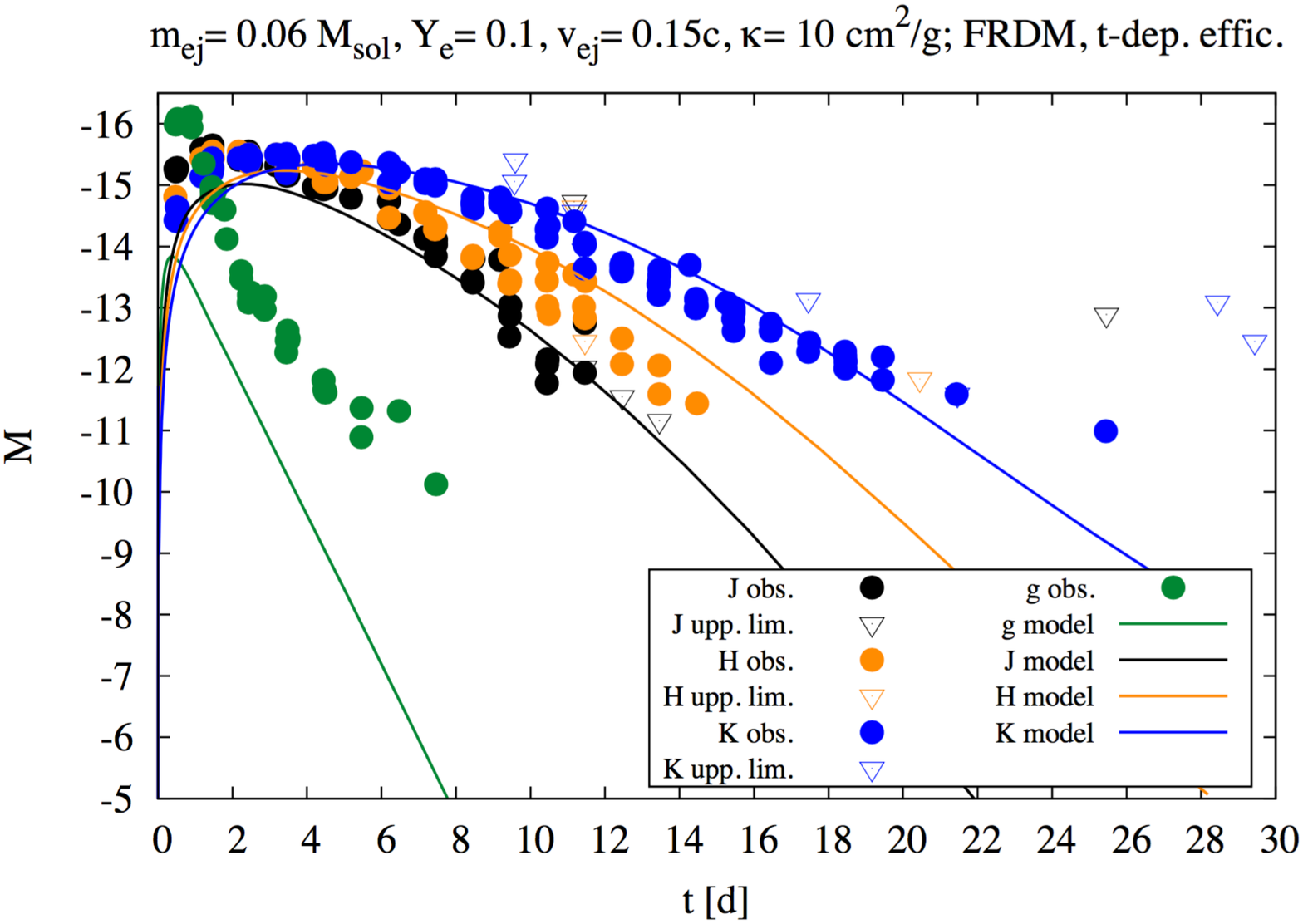}
\includegraphics[width=1.\columnwidth,angle=0]{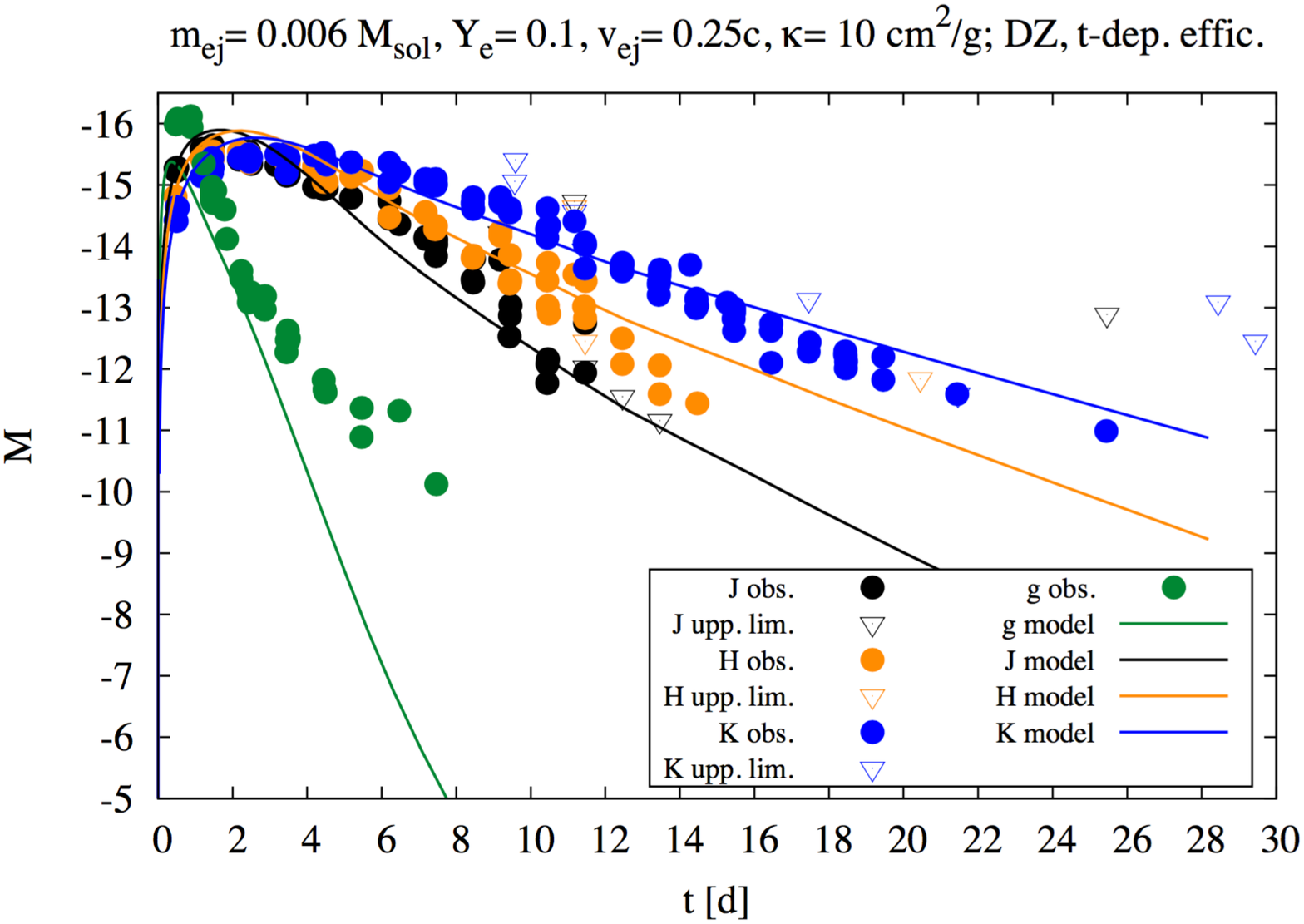}
\caption{Comparison of a low-$Y_e$ (= 0.1) matter case, 
representative for a "tidal" dynamical ejecta, with NIR $JHK$-band 
observations (https://kilonova.space/kne/GW170817/). 
The detailed ejecta model parameters are shown at the top of each panel.
The left panel uses heating according to the FRDM nuclear mass model, for the right panel
a DZ-type heating rate has been employed (see Sec. 3 in the main text for a discussion).
}
\label{fig:magnitudes_model_obs_WITH}
\end{figure*}
\begin{figure}
\centerline{
\includegraphics[width=1.\columnwidth,angle=0]{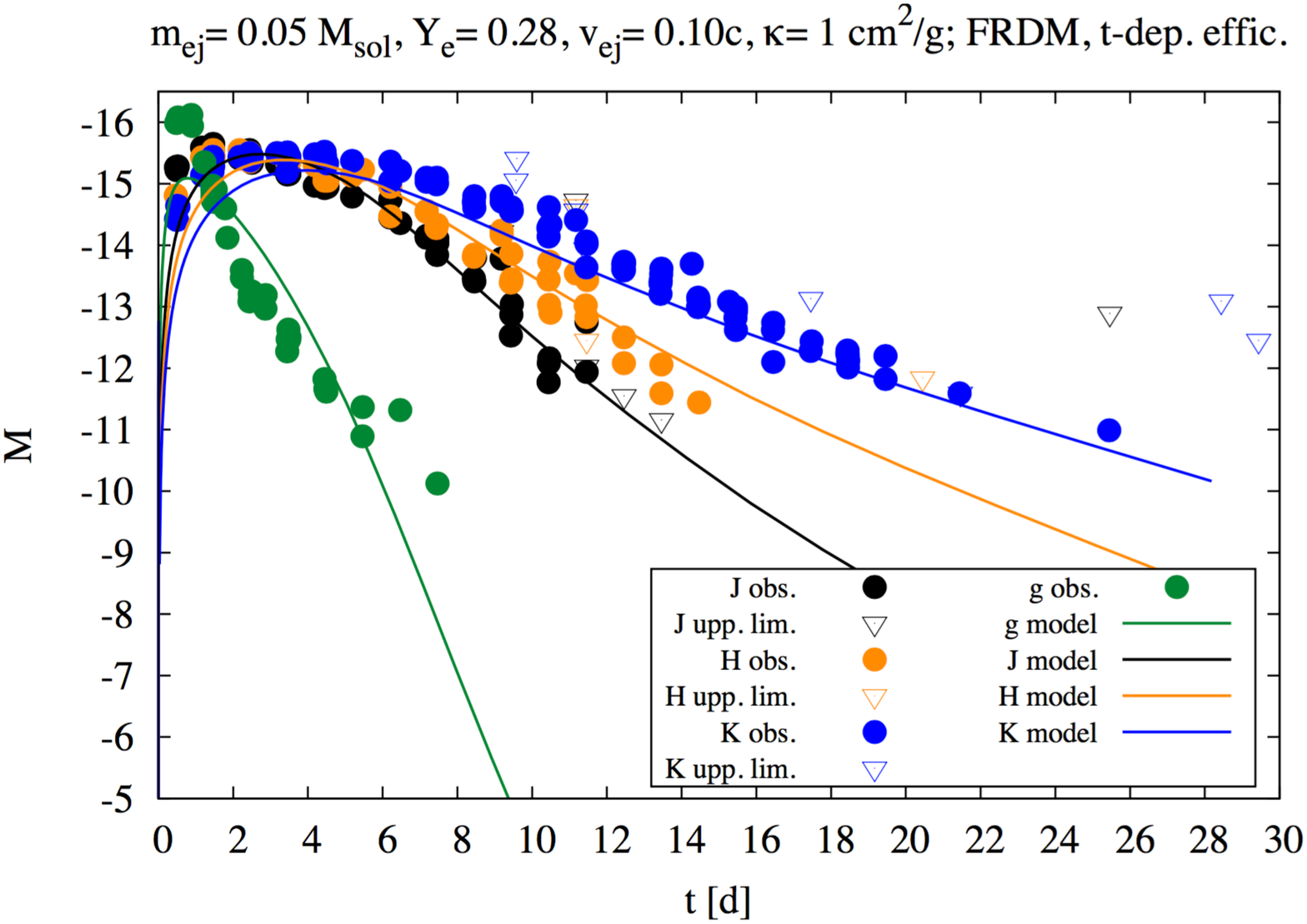}
}
\caption{ 
Comparison for a model {\em without} lanthanides and third r-process peak ejecta. 
The detailed ejecta model parameters are shown at the top of the panel. For this 
calculation the heating rate of the FRDM nuclear mass model was used
(see Sec. 3, main text for a discussion).
The properties of this model are characteristic for matter unbound from
an accretion torus.}
\label{fig:magnitudes_model_obs_WITHOUT}
\end{figure}

The most conservative expectation prior to GW/EM170817 was
a red EM-transient due to high-opacity ejecta peaking days
after the GW-chirp \citep{kasen13a,tanaka13a,barnes16a,rosswog17a,wollaeger17a}.
Although the emergence of an additional blue component was discussed in
theoretical work
\citep{barnes13a,rosswog14a,grossman14a,metzger14,perego14b,martin15,fernandez15,kasen15a},
the brightness of the blue optical transient (AT2017gfo) that was 
detected \citep{coulter17} hours after the GW-chirp came as 
a surprise to most in the community. It  can be 
explained by lower opacity material, potentially accelerated to mildly relativistic
velocities through a GRB-jet producing a cocoon while plowing through previously
ejected material \citep{kasliwal17} or by a strong wind with moderately high electron fraction
\citep{evans17}. This early blue component will not be discussed here.\\
Instead, we focus on the late NIR emission in the J-, H- and K-bands. We have
explored the parameter space in electron fraction, velocity, and ejected mass in 
more than 220 nuclear network based macronova simulations.
For each model the initial conditions are set up as described in detail in Sec. 2.4.3
of \cite{rosswog17a}) and the nuclear heating history $\dot{q}(t)$ is calculated using the
WinNet reaction network with the FRDM mass formula. We use time-dependent heating 
efficiencies $f_{\rm tot}$ based on the work of \cite{barnes16a} as calculated in
\cite{rosswog17a}. Here we use the time-dependent averages of the FRDM-cases
explored in the latter work (their Fig. 8).
We account for the uncertainty in the nuclear heating
rate due to the $\alpha-$decay of trans-lead nuclei (as discussed 
in Sec.~\ref{sec:heating}) in some experiments by enhancing the net heating 
rate of the FRDM results by a factor of 5 and refer to it as "DZ-type heating".\\
To extract the radiative signature we use a semianalytic eigenmode expansion formalism
based on \cite{pinto00a}. This semi-analytic approach has been shown to yield
good agreement with more complex radiative transfer models and represents 
an improvement over the simpler model of \cite{grossman14a} that we used in 
earlier work. Our approach is briefly summarized  in Appendix A.\\
The NIR lightcurves alone leave some ambiguity as to what the exact
ejecta parameters are, but they can be significantly constrained further
if more data sets are taken into account.
Interesting examples of NIR lightcurves
are shown in Fig.~\ref{fig:magnitudes_model_obs_WITH}.
The low $Y_e$  is characteristic for the  "tidal" component of 
dynamic ejecta that is ejected immediately during the merger at its original, 
very low electron fraction and produces substantial r-process contributions 
from $A\approx 100$ up to and  beyond the platinum peak. The left panel
shows a good fit of the NIR light curves for the case that the heating rate
from the FRDM nuclear mass model is used ($m_{\rm ej}= 0.06$ \msun, 
$Y_e= 0.1$, $v_{\rm ej}= 0.15c$, $\kappa=10$ cm$^2$/g).
If instead  DZ-type nuclear heating is used, our best parameters differ
from the FRDM case ($m_{\rm ej}= 0.006$ \msun, 
$Y_e= 0.1$, $v_{\rm ej}= 0.15c$, $\kappa=10$ cm$^2$/g), and in 
particular substantially less mass is required.\\
Interestingly, the NIR late-time light curves do not
necessarily prove the presence of either lanthanides or 
third r-process peak elements, although based on theoretical 
modelling their presence is certainly expected. It is also possible
to obtain a good fit for an electron fraction ($Y_e= 0.28$) that is large enough 
to avoid the production of lanthanides and the 
third r-process peak and thus has a lower effective 
opacity ($\kappa=1$ cm$^2$/g),
see Fig.~\ref{fig:magnitudes_model_obs_WITHOUT}. 
The mass of $0.05$ \Msun could plausibly be ejected from a 
$\approx 0.13$ \Msun torus (assuming 40\% ejection) and also the electron
fraction is in the range expected for matter that has been exposed to a merger
background neutrino field \citep{qian96b,rosswog14a,perego14b,siegel17a}. Only the 
velocities are larger (by a factor of $\sim 2$) than what simulations \citep{fernandez13a,just15} 
have found so far for unbound torus matter.\\
In order to see whether the comparison with another band can break the
degeneracy between matter with and without lanthanides, we have added  
the g band (green) to Figs.~\ref{fig:magnitudes_model_obs_WITH} 
and \ref{fig:magnitudes_model_obs_WITHOUT}. Based on this comparison alone,
there would be a slight advantage for the lanthanide-free case. We would, however,
consider it very unlikely that a merger starting out from cold, high-density $\beta$-equilibrium
with $Y_e\approx 0.06$ manages to raise the electron fractions of {\em all} the ejecta beyond
the critical value of $\approx 0.25$. To conclusively decide between the two cases may
be beyond the capabilities of the current modelling (ours and in general).
An obvious caveat is our use of constant gray opacities. In reality, 
opacities and the position of the photosphere are wavelength dependent. 
Another strong limitation stems from using only one value for electron fraction and velocity.
This clearly is a strong simplifications and what has been observed is 
a superposition of distributions of physical conditions.

\section{Discussion}
\begin{figure}
\vspace*{0.cm}
\centerline{\includegraphics[width=1\columnwidth,angle=0]{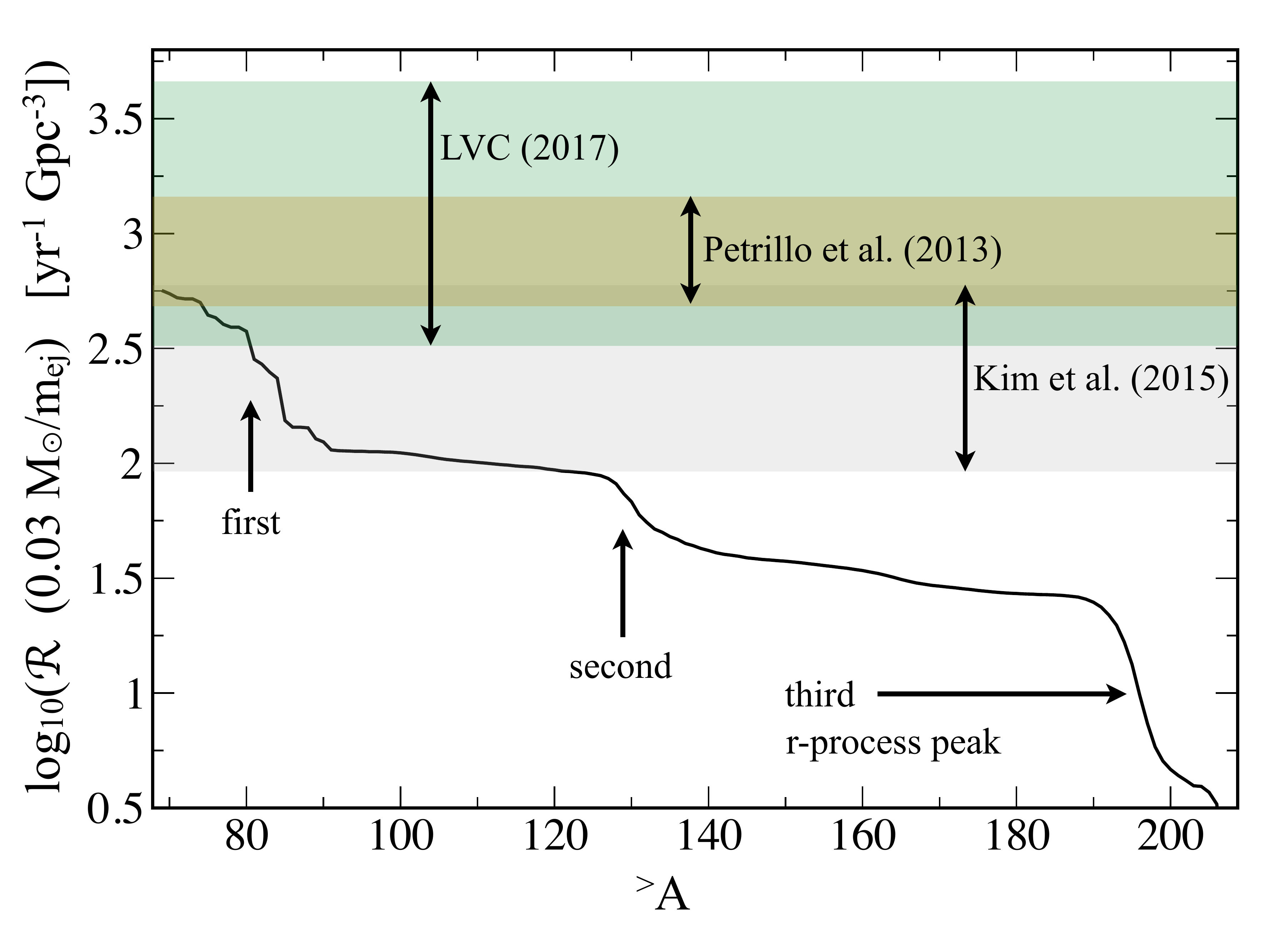}}
\vspace*{0cm}
\caption{Needed event rates, scaled to an ejecta mass of 0.03 \msun, if NSNS mergers are to produce all r-process (in solar
proportions) {\em above a minimum nucleon number} $^>$A (solid black line). Also shown are the estimated rates (90\% conf.) for NSNS mergers from the population synthesis study of \cite{kim15}, the sGRB rates based on SWIFT data from \cite{petrillo13} and the LVC estimate based on the first detected NSNS merger event. 
}
\label{fig:needed_rates}
\end{figure}
\begin{figure}
\vspace*{0.cm}
\centerline{\includegraphics[width=1.02\columnwidth,angle=0]{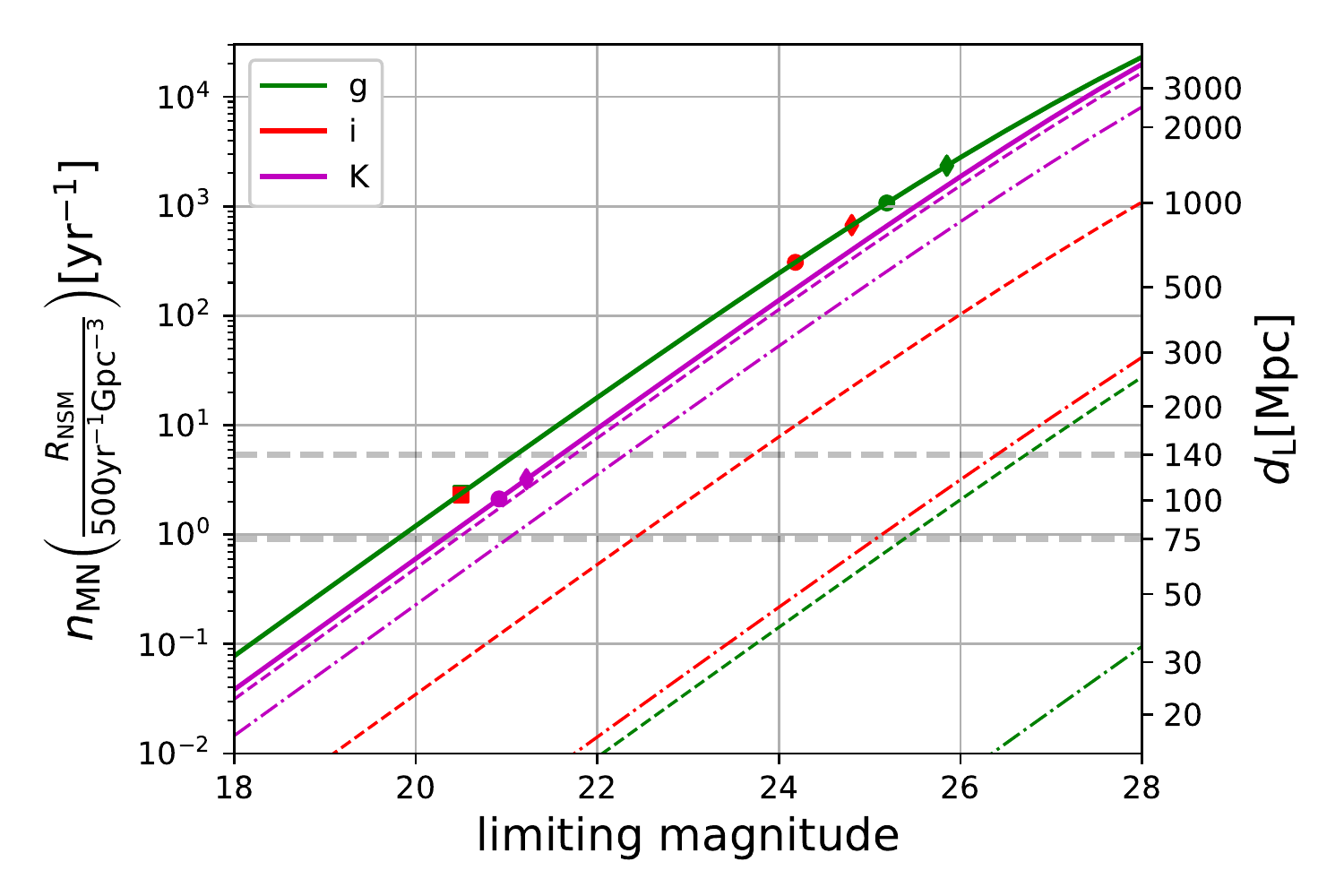}}
\caption{Expected number of transients similar to AT2017gfo that peak above a given g-, i- or K-band limiting magnitude. Calculations are based on the black-body model presented in \cite{kasliwal17}. Solid lines are based on the peak brightness, while dashed and dash-dotted lines are based on the brightness at 4 and 7 days after the merger, respectively. Note that the result for i-band at peak is not shown because it is practically the same as for g-band. The circles and diamonds correspond to the depths of 60- and 180-second exposures, respectively (for LSST in g- and i-band and for VISTA in K-band). The square marker shows the depth of a ZTF 600-second exposure in the g-band. The dashed lines show assumed ranges for GW detections of 75~Mpc for NSNS mergers and 140 Mpc for NS-BH.   
}
\label{fig:rate-bb}
\end{figure}

The observation of GW170817 is a milestone. The first direct observation of a neutron star merger
and its coincident electromagnetic detection has finally proven two long-held suspicions, namely
i) that such mergers are a source of short GRBs\footnote{It is currently debated whether the event was
a typical short GRB. While \cite{kasliwal17} argue that this was not a classical short burst GRB seen off-axis
and interpret the emission instead in terms of a cocoon model, \cite{lyman18} find the late optical emission
being consistent with the expectations from a structured jet that would --if seen on-axis-- have been 
interpreted as a high-luminosity short GRB.}
and --as we have demonstrated here-- 
ii) it provides a first {\em direct} proof that their ejecta are a major source for the cosmic r-process 
nucleosynthesis.\\
We have explored the radioactive heating rate for a broad range of physical conditions
and we find that the decline of the observed bolometric luminosity of AT2017gfo agrees
very well with the decay produced by matter with $Y_e \lesssim 0.3$, but not larger. 
Such matter is subject to the rapid-neutron capture process, see Fig.~\ref{fig:nucleo_all_yes}.
The bolometric lightcurve rules out in particular nickel winds as the major source of the 
emission.
This provides strong, {\em direct} observational evidence for neutron star mergers being a 
major nucleosynthesis site and confirms earlier purely theoretical or indirect 
observational conclusions \citep{lattimer74,rosswog98b,rosswog99,freiburghaus99b,korobkin12a,hotokezaka15a,beniamini16a}.\\
Using nuclear network calculations employing the FRDM nuclear mass model, we 
derive a lower limit on the ejecta mass of $\approx 1.5 \times 10^{-2}$ \Msun to 
explain the bolometric luminosity. Due to uncertainties in the nuclear physics far from stability, this limit could potentially be reduced by a factor of up 
to $\sim 5$. Even in this most pessimistic case
the real ejecta amount would likely be $\sim 1\%$ of a solar mass,
which is a substantial amount in a cosmic nucleosynthesis context.
Based on this first detected GW-event, the NSM rate (90\% conf.) is 
estimated as 320 - 4740 Gpc$^{-3}$ yr$^{-1}$ \citep{LIGO17_GW_NSNS},
compact object merger rate estimates based on SWIFT sGRB data point to
$\sim 500 - 1500$ Gpc$^{-3}$ yr$^{-1}$ \citep{petrillo13} while recent population synthesis studies \citep{kim15}
estimate the rate\footnote{We use the density of Milky Way equivalent galaxies of \cite{abadie10} to transform
between different units.} as $244^{+325}_{-162}$  Gpc$^{-3}$ yr$^{-1}$,
which means that within the rate uncertainties, neutron star mergers can well 
produce all the r-process elements in the MW 
($M_r\sim 19 \; 000$ \Msun; \citealt[e.g.][]{bauswein14b,shen15a,rosswog17a}),
\be
M_r \sim 17 \; 000 {\rm M}_\odot \; \left(\frac{\mathcal{R}_{\rm NSNS}}{500 \rm Gpc^{-3} \; yr^{-1}} \right)\left(\frac{\bar{m}_{\rm ej}}{0.03 M_\odot}\right) \left(\frac{\tau_{\rm gal}}{1.3 \times 10^{10} {\rm yr}}\right).
\ee
Clearly, which rate is needed depends on which r-process
elements are produced. In Fig.~\ref{fig:needed_rates},
we show as solid black line the required event rate 
(scaled to an ejecta mass of 0.03 \msun) under
the assumption that NSMs produce all r-process 
(in solar proportions) above a limiting nucleon number $^>$A. 
So if all r-process is produced in NSMs, an event rate of about
560 (0.03 \msun/$\bar{m}_{\rm ej}$) yr$^{-1}$ Gpc$^{-3}$ is needed. 
If instead, NSMs should only produce r-process beyond
the second peak ($A>130$), a rate of only 70 
(0.03 \msun/$\bar{m}_{\rm ej}$) yr$^{-1}$ Gpc$^{-3}$ would suffice.
The early blue emission observed in AT2017gfo, however, 
is most naturally explained with lower-opacity ejecta and 
therefore argues for the production of at least some
lower-mass r-process material, which would also be consistent 
with recent theoretical studies
\citep{wanajo14,perego14b,just15,wu16}. This could 
point to rates between the above two extremes. 
From the modelling of the NIR lightcurves alone it is not
possible to distinguish between a pure high-opacity and pure 
low-opacity case (Figs.~\ref{fig:magnitudes_model_obs_WITH} and ~\ref{fig:magnitudes_model_obs_WITHOUT}), but
merger simulations indicate that at least some low-$Y_e$ 
matter is ejected and this is also consistent with the broad 
spectral features that have been observed \citep{kasliwal17,tanvir17,chornock17}. 
Therefore, we interpret this first event as strong evidence
for a broad range of r-process nuclei being produced and not
just --as thought until a few years ago-- only $A>130$ 
material.\\
Based on the discussed numbers, NSMs could produce all the cosmic r-process
without needing an additional production site. But within the uncertainties of
rates/ejecta masses additional contributors are certainly possible.
It has been argued (see e.g. the discussion in \citet{thielemann17}) that an additional source of strong r-process would make
it easier to explain the very large scatter of [Eu/Fe] observed already at very low
metallicities and the presence of so-called "actinide boost" stars \citep{lai08} that have a
$\sim$ fourfold enhancement of thorium and uranium relative to europium. As a potential additional strong r-process source
core-collapse supernovae have been suggested \citep{winteler12b} that eject
r-process matter in magnetohydrodynamial jets. Recent 3D MHD studies 
\citep{moesta17}, however, find that such jets are subject to instabilities
unless the initial star is endowed with a (likely unrealistically large) pre-collapse field
of $\sim10^{13}$ G. If such instabilities set in, matter is exposed for longer
to the central neutrino emission and therefore raises its $Y_e$ to large enough
values to avoid significant platinum peak contributions. This interesting topic 
certainly warrants more work in the future. For now we conclude, that additional
contributions at both the light and heavy r-process end are possible, but are
--based on current numbers-- not strictly required. It is fair to state, 
however, that the corresponding chemical galactic evolution questions are not yet
fully understood and will require further studies.
\\
Using the same method as described in \citet[][their Sect. 3.4]{rosswog17a}, we 
estimate the expected number of events like AT2017gfo that peak above a 
given limiting magnitude, see Fig.~\ref{fig:rate-bb}. For this, we used 
the blackbody model described in \cite{kasliwal17} and a reference event 
rate of 500 Gpc$^{-3}$~yr$^{-1}$. Due to the early blue peak of the 
observed transient, the expected numbers in the optical are large.
A survey like ZTF ($g<22$~mag 600-second exposures for GW follow up) 
could detect all NSMs with such a blue peak within the LIGO range, 
of which we would expect approximately one per year. With a larger optical survey 
telescope such as LSST $\sim1000$ macronovae could become observable per year. This, however, 
requires that the follow-up is triggered the same night because g-band fades rapidly and
the numbers drop to only one event per year with 4 days after the merger. 
Observations at longer wavelengths would provide a larger window. In i-band, the number of 
observable macronovae 4 days after the merger is nearly two orders of magnitude larger. 
Similarly in the NIR, a 60-second exposure with VIRCAM in the K band would be sufficient 
and the transient remains observable for more than a week. 
\\
The long awaited era of multi-messenger GW-astronomy has now finally begun 
and the first multi-messenger detection of a merging neutron star binary 
has conclusively proven the long-held conjectures of producing
short GRBs and forging heavy elements, thereby providing the first 
{\em directly} observed constraints of rates and ejecta masses. How 
representative this first event was will have to be probed by 
future multi-messenger detections.\\

\begin{acknowledgements}
      We thank the anonymous referee for her/his insightful comments that helped
      the improve our paper. It is a great pleasure to thank Friedrich-Karl Thielemann
      for insightful discussions.
      SR has been supported by the Swedish Research Council (VR) under grant number 2016- 03657\_3, 
by the Swedish National Space Board under grant number Dnr. 107/16. SR, JS and AG are supported
by the research environment  grant "Gravitational Radiation and Electromagnetic Astrophysical Transients 
(GREAT)" funded by the Swedish Research council (VR) under Dnr 2016- 06012. 
MMK acknowledges support by the GROWTH project funded by the National Science Foundation under PIRE Grant No 1545949.
This work has further been supported by  the CompStar network, COST Action MP1304. 
Some of the simulations were performed on the resources provided by  the North-German 
Supercomputing Alliance (HLRN). A portion of this work was also carried out under the auspices 
of the National Nuclear Security Administration of the 
U.S. Department of Energy at Los Alamos National Laboratory 
under Contract No. DE-AC52-06NA25396 (O.K., R.W.). Some of
the simulations used in this work were performed on the resources
provided by the Los Alamos National Laboratory Institutional 
Computing Program (O.K.,R.W.). OK and RW are thankful to Aimee L. Hungerford, Chris L. Fryer 
and Christopher J. Fontes for inspiring and productive discussions.
\end{acknowledgements}

\begin{appendix}
\section{Summary of the semi-analytic model}
Here we briefly summarize the major ingredients of
our semi-analytic macronova model. It uses the analytic
density structure found from solving the spherical Euler
equations for a self-similar homologous flow that is
derived in Sec. 2.1.1 of \cite{wollaeger17a}. To extract
the radiative signature, it makes use of the analytic
solution of the comoving frame transport equations for
constant opacity as derived by \citet{pinto00a} in the 
context of type Ia supernovae. 
This solution has been carefully cross-checked against
a multigroup radiative transfer code, see Sec.2.3.1
of \cite{wollaeger17a}.\\
%
%
The heating rates $\dot{q}$ 
that enter the model can be chosen to be either a) an analytic
prescription, as Eq.(4) in \cite{korobkin12a} or b) be taken as the output
of a nuclear reaction network calculation. In all 
calculations shown in this paper we use the output
of a nuclear reation network calculation performed with
WinNet \citep{winteler12,winteler12b} and the FRDM mass
model. For the heating efficiencies $f_{\rm tot}$ 
we use the time-dependent averages of all the FRDM models
shown \cite{rosswog17a}, which, in turn, are based on
the work of \cite{barnes16a}.
As discussed in the main text, Sec. 2, we mimick the results
of the Duflo-Zuker mass formula by multiplying the net heating
rates ($\dot{q} \; f_{\rm tot}$) by a factor of 5 for 
those cases where nuclei heavier than lead are formed.\\
%
%
For our expansion model we follow \cite{wollaeger17a} and 
start from the mass and momentum
equations of ideal, non-relativistic hydrodynamics for
a sphere that expands into vacuum. For self-similar homologous 
flow a scale parameter $R(t)$ and a shape function $\varphi(x)$
can be introduced, $x$ being the dimensionless radial coordinate 
$x= r/R(t)$, so that density and velocity are
\begin{equation}
\rho(t,r)= \frac{\varphi(x)}{R(t)^3} \quad {\rm and} \quad
 v(t,r)  = r \frac{\dot{R}(t)}{R(t)}.
\end{equation}
If a polytropic equation of state with $\Gamma= 4/3$ 
(``radiation dominated flow'') is used, the Euler equations 
admit a closed form solution:
\begin{equation}
\varphi(x)= \rho_0 R_0^3 \; (1-x^2)^3
\end{equation}
and 
\begin{equation}
(t-t_0)= \frac{R(t)}{v_{\rm max}}\sqrt{1- \frac{R_0}{R(t)}}+ \frac{R_0}{v_{\rm max}} \log\left[\frac{R(t)}{R_0} \left( 1- \sqrt{1-\frac{R_0}{R(t)}}\right)^2 \right],\label{eq:R_nonlin}
\end{equation}
where $R_0= R(t_0)$, $\rho_0$ the initial central density and $v_{\rm max}$ the
expansion velocity. For $t \gg t_0$ Eq.(\ref{eq:R_nonlin}) reduces to $R(t)\simeq v_{\rm max} t$ and the density profile becomes
\begin{equation}
\rho(t,r)= \rho_0 \left(\frac{t_0}{t}\right)^3 \left(1 - \frac{r^2}{v_{\rm max}^2 t^2} \right)^3,
\end{equation}
where $v_{\rm max}$ is the expansion front velocity. The ejecta mass and 
average velocity are
\begin{equation}
m_{\rm ej}= \frac{64 \pi}{315}\rho_0 t_0^3 v_{\rm max}^3 \quad {\rm and} 
\quad \bar{v}= \frac{63}{128} v_{\rm max}.
\end{equation}

%
%
Coming from a radioactively heated, initially optically thick cloud
of matter, macronova emission bears some similarity with type Ia
supernovae. Since the energy injection decreases rapidly with time, 
the light curve will peak as soon as the injected energy has a chance
to escape being converted into kinetic energy. This happens when the
diffusion time becomes comparable to the elapsed time.
In our model, we extract the radiative signature based on an
eigenmode expansion formalism developed by \citet{pinto00a}. 
Our notation and details of derivation follow \cite{wollaeger17a},
where it was applied in the context of macronova with uniform density (see
their Appendix A).
Here we only summarize the main points, see 
the original paper for the justification of the physical assumptions
and for the detailed derivation.
\\
The starting point is the semi-relativistic diffusion equation
which for radiation dominated flows reads
\begin{equation}
\frac{DE}{Dt}
 - \nabla\cdot\left(\frac{c}{3\kappa\rho}\nabla E\right)
 + \frac43 E\nabla\cdot\vec{v}
 = \rho\dot{q}(t),
\label{eq:basic}
\end{equation}
where $E$ is the internal energy density, $D/Dt$ the Lagrangian 
time derivative and $\kappa$ the constant (gray) opacity. Using 
dimensionless quantities and assuming spherical symmetry and 
homologous expansion one finds
\begin{align}
\frac{DE}{Dt}
 - \frac1{R^2 x^2}
   \left[\frac{c}{3\kappa\rho} x^2 E'\right]'
 + \frac{4E}{t}
 = \rho\dot{q}(t),
\label{eq:basicx}
\end{align}
where the primed quantities are being differentiated with respect to $x$.
The second term in (\ref{eq:basic}) is the divergence of the radiative
diffusion flux $\vec{F}$:
\begin{align}
\vec{F} = -\frac{c}{3\kappa\rho} \nabla E
   \equiv -\frac{c}{3\chi} \nabla E,
\label{eq:flux}
\end{align}
where we introducted the extinction coefficient $\chi\equiv\kappa\rho$.
%
%
With a subsequent separation of variables in mind, we make the Ansatz
\begin{align}
  E(x,t) = E_0 \left[\frac{t_0}{t}\right]^4\psi(x)\phi(t), \quad {\rm and} \quad
  \rho(x,t) = \rho_0 \left[\frac{t_0}{t}\right]^3\ji(x).
\label{eq:ansatz}
\end{align}
and recast the equation (\ref{eq:basicx}) into the following form:
\begin{align}
\frac{t_0}{t}\psi\dot\phi
- \phi
  \frac1{\tau_0 x^2}\left[\frac{x^2\psi'}{\ji}\right]'
= \frac{\rho_0 \dot{q}(t)}{E_0}\ji,
\end{align}
where we introduced the timescale
$\tau_0\equiv\frac{3\kappa\rho_0 R_0^2}{c}$.
\\
It is convenient to use the rescaled time coordinate $\zeta\equiv t/t_0$ with
$\bar\tau_0\equiv\tau/t_0$:
\begin{align}
\psi\frac{\pd\phi}{\pd\zeta}
- \zeta\phi
  \frac1{\bar\tau_0 x^2}\left[\frac{x^2\psi'}{\ji}\right]'
= \frac{\rho_0 t_0 \dot{q}(\zeta)}{E_0}\ji.
\label{eq:inhom}
\end{align}
The corresponding homogeneous linear equation,
\begin{align}
\frac1{\zeta\phi(\zeta)}\frac{\pd\phi(\zeta)}{\pd\zeta}
- \frac1{\bar\tau_0 \psi(x) x^2}\left[\frac{x^2\psi(x)'}{\ji(x)}\right]'
= 0,
\end{align}
admits a separation of variables for some constant $\lambda$:
\begin{align}
&\frac{\bar\tau_0}{\zeta\phi(\zeta)}\frac{\pd\phi(\zeta)}{\pd\zeta}
= -\lambda,
\label{eq:tlam}
\\
&\left[\frac{x^2 \psi(x)'}{\ji(x)}\right]'
+ \lambda x^2\psi(x) = 0.
\label{eq:eigen}
\end{align}

Equation (\ref{eq:eigen}) is an eigenvalue problem.
We can now make a substitution $\psi(x)\to(1-x^2)^4z(x)$ to regularize it
at the outer boundary, where density becomes zero. The power 4 is
motivated by the following reasoning: for an adiabatic radiation-dominated
outflow with a constant entropy $T^3/\rho={\rm const.}$
the temperature profile should be $\propto(1 - x^2)$,
provided that the density profile is $\propto(1-x^2)^3$.
The correspondingly the internal energy density
$E\propto T^4\propto(1-x^2)^4$.

The eigenvalue problem can be cast into Sturm-Liouville form:
\begin{align}
\frac{d}{dx}\left[x^2(1-x^2)^5\frac{dz}{dx}\right]
+ x^2(1-x^2)^4\left[\lambda(1-x^2)^4-24\right]z = 0,
\label{eq:sturm}
\end{align}
for which there exists a spectrum of distinct real eigenvalues 
$\left\{\lambda_m\right\}$ and an orthogonal basis $\left\{z_m(x)\right\}$ 
in Hilbert space with respect to the scalar product
\begin{align}
\langle f|g\rangle
\equiv
\int_0^1 x^2(1-x^2)^8 f(x) g(x) dx.
\end{align}

This is a well-posed eigenvalue problem which can be solved using a variety of
numerical methods. We use a Galerkin method with linear finite element discretization
on a uniform grid with $N=100$ points.
%
%
Having computed the eigenvalues and eigenfunctions, we can expand a solution 
to the inhomogeneous problem (\ref{eq:inhom}) in eigenfunctions with
time-dependent expansion coefficients $\phi_m(\zeta)$:
\begin{align}
  E(t,x)= E_0 \zeta^{-4}\sum_m \phi_m(\zeta)\psi_m(x),
  \label{eq:inan}
\end{align}
where the functions $\psi_m(x)\equiv(1-x^2)^4 z_m(x)$ are weighted
eigenfunctions of Eq.(\ref{eq:sturm}). 
Equation (\ref{eq:tlam}) splits into a series of decoupled first-order ODEs
for the functions $\phi_m(\zeta)$:
\begin{align}
  \left(\frac{d\phi_m}{d\zeta} +
        \frac{\zeta}{\bar\tau_0}\phi_m\lambda_m
  \right) N^2_m
  =
  \frac{\rho_0 t_0\dot{q}(\zeta)}{E_0}\zeta d_m,
\end{align}
where
\be
  N^2_m  \equiv \int_0^1 x^2\psi^2_m(x) dx
         \equiv \int_0^1 x^2(1-x^2)^8 z^2_m(x) dx,
\ee
and
\be
  d_m    \equiv \int_0^1 x^2\ji(x)\psi_m(x) dx
         \equiv \int_0^1 x^2(1-x^2)^7z_m(x) dx.
\ee
We solve these equations numerically with a Crank-Nicholson
integrator.
\\
The bolometric luminosity is proportional to the flux~(\ref{eq:flux}) 
through the outer boundary at $x=1$:
\begin{align}
L(t) &= 4\pi R(t)^2 
      \cdot\left.
           \left(-\frac{c}{3\kappa\rho}\cdot\frac{\pd E}{\pd r}\right)
           \right|_{x=1}\\
  &=-\frac{4\pi c R_0 E_0}{3\kappa\rho} 
     \left.
     \sum_{m=1}^N \phi_m\left(\frac{t}{t_0}\right) \psi'_{m}(x)
     \right|_{x=1}\\
  &= \frac{32\pi c R_0 E_0}{3\kappa\rho_0} 
     \sum_{m=1}^N \phi_m\left(\frac{t}{t_0}\right) 
     \cdot \left.z_m(x)\right|_{x=1},
\end{align}
where we used the following simplification to express the derivative at $x=1$:
\begin{align}
  \left.\frac{\psi'_m(x)}{\rho(x)}\right|_{x=1} 
  = \left.\frac{\left[(1-x^2)^4 z_m(x)\right]'}{\rho_0(1-x^2)^3}\right|_{x=1} 
  = -8 \left.z_m(x)\right|_{x=1}.
\end{align}
%
%
\begin{figure*}
\begin{center}
\begin{tabular}{c}
\includegraphics[width=0.49\textwidth]{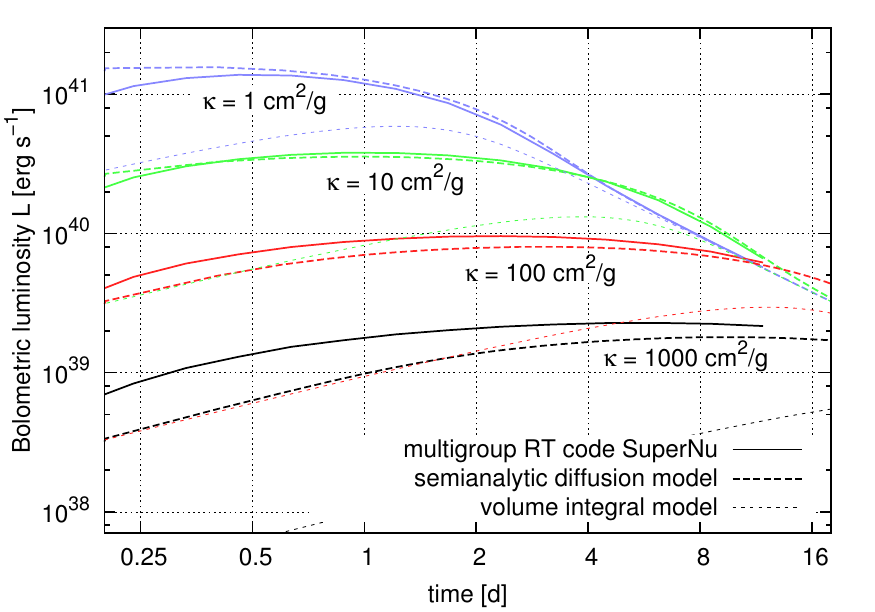}
\includegraphics[width=0.49\textwidth]{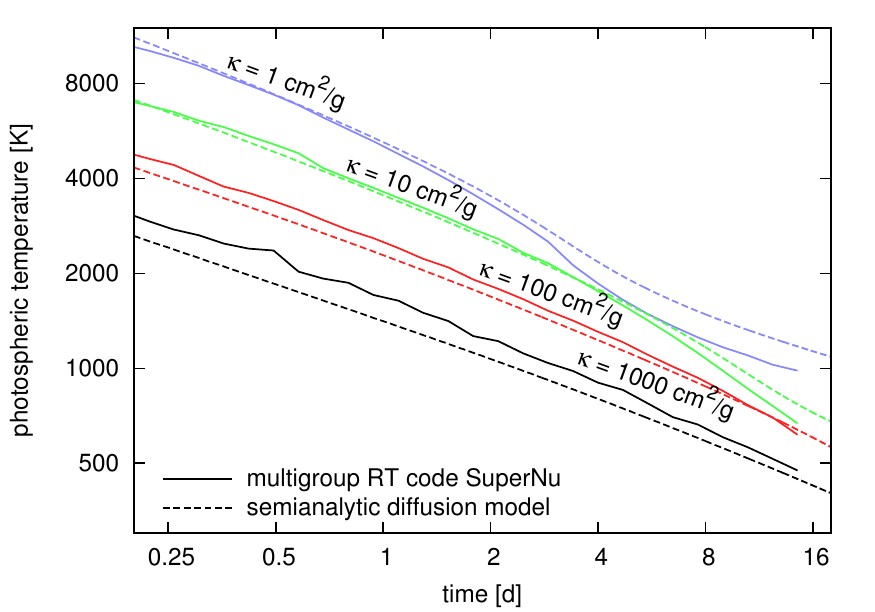}
\end{tabular}
\end{center}
\caption{Left: Bolometric luminosity for the described semianalytic diffusion model 
(dashed lines) with opacities 
$\kappa = 1, 10, 100, 1000\ {\rm cm}^2$/g compared
against full multigroup Monte Carlo radiative transfer models
(solid lines) and the substantially simpler model of 
Grossman et al. (2014) that uses volume integration over the radiative
zone ('volume integral model'). Right: Comparison of the photospheric 
temperature evolution between our semianalytic diffusion model (dashed)
with the radiative transfer code SuperNu (solid).}
\label{fig:compare-supernu}
\end{figure*}
To scrutinize our approach, we have computed light curves 
for a range of opacities 
$\kappa = 1, 10, 100, 1000\ {\rm cm}^2$/g and a power-law heating
rate $\dot{q}(t)=q_0 t_d^{\alpha}$ 
with $q_0 = 5\times10^9\ {\rm erg}\ {\rm g}^{-1}{\rm s}^{-1}$ and $\alpha= -1.3$.
Fig.~\ref{fig:compare-supernu} presents the comparison of bolometric light
curves caluclated with the described approach and the corresponding one
from the full multigroup Monte Carlo radiative transfer code 
{\tt SuperNu}\footnote{
\url{https://bitbucket.org/drrossum/supernu/wiki/Home}}. 
The semianalytic diffusion model performs substantially better than 
the simpler analytic solution of \cite{grossman14a}, which uses integration of
the energy rate over the radiative zone outside of the trapped region.
\\
As demonstrated in~\cite{wollaeger17a}, the spectrum of gray opacity models is
well approximated by a blackbody with effective temperature at the
photosphere:
\begin{align}
T_{\rm eff}(t) = \sqrt[\leftroot{5}\uproot{2}4]
                      {\frac{L(t)}{4\pi\sigma R_{\rm ph}(t)^2}}.
\label{eq:teff}
\end{align}
The right panel in ~\ref{fig:compare-supernu} shows the effective photospheric
temperature evolution for full radiative transfer models and the temperatures
for semianalytic models, computed using expression (\ref{eq:teff}) and
assuming the photosphere at optical depth $\tau_{\rm ph}=2/3$.

\end{appendix}


\begin{thebibliography}{}

 \bibitem[\protect\citeauthoryear{{Abbott}, {Abbott}, {Abbott} \&
  {Acernese}}{{Abbott} et~al.}{2017a}]{abbott17_MM}
{Abbott} B.P. et al. 2017, ApJ, 848, L12
  
\bibitem[\protect\citeauthoryear{Abbott et al.}{2017b}{}]{abbott17b}
Abbott et al. 2017, Nature, 551, 85

\bibitem[\protect\citeauthoryear{Abbot et al.}{2017b}{}]{ligo_NSNS_MultiMessenger17}
Abbot, B. P. et al. 2017, ApJL 848, 2 

\bibitem[\protect\citeauthoryear{{Abadie}, {Abbott}, {Abbott}, {Abernathy},
  {Accadia}, {Acernese}, {Adams}, {Adhikari}, {Ajith}, {Allen} \& et
  al.}{{Abadie} et~al.}{2010}]{abadie10}
{Abadie} J.,  {Abbott} B.~P.,  {Abbott} R.,  {Abernathy} M.,  {Accadia} T.,
  {Acernese} F.,  {Adams} C.,  {Adhikari} R.,  {Ajith} P.,  {Allen} B.,    et
  al.  2010, Classical and Quantum Gravity, 27, 173001
  
\bibitem[Arcavi et al.(2017)]{arcavi17} Arcavi, I., Hosseinzadeh, G., Howell, D.~A., et al.\ 2017, Nature, 551, 64 
  
\bibitem[\protect\citeauthoryear{{Argast}, {Samland}, {Thielemann} \&
  {Qian}}{{Argast} et~al.}{2004}]{argast04}
{Argast} D.,  {Samland} M.,  {Thielemann} F.-K.,    {Qian} Y.-Z.  2004, A\&A,
  416, 997
  
\bibitem[\protect\citeauthoryear{{Barnes} \& {Kasen}}{{Barnes} et~al.}{2013}]{barnes13a}
{Barnes} J.,  {Kasen} D.  2013, ApJ, 775, 18 

\bibitem[\protect\citeauthoryear{{Barnes}, {Kasen}, {Wu} \&
  {Martinez-Pinedo}}{{Barnes} et~al.}{2016}]{barnes16a}
{Barnes} J.,  {Kasen} D.,  {Wu} M.-R.,    {Martinez-Pinedo} G.  2016, ApJ,
  829, 110

\bibitem[\protect\citeauthoryear{{Bauswein}, {Ardevol Pulpillo}, {Janka} \&
  {Goriely}}{{Bauswein} et~al.}{2014}]{bauswein14b}
{Bauswein} A.,  {Ardevol Pulpillo} R.,  {Janka} H.-T.,    {Goriely} S. 2014,
  ApJL, 795, L9

\bibitem[\protect\citeauthoryear{{Beloborodov}}{{Beloborodov}}{2008}]{beloborodov08}
{Beloborodov} A.~M.,  2008, in {M.~Axelsson} ed., American Institute of Physics
  Conference Series Vol.~1054 of American Institute of Physics Conference
  Series, {Hyper-accreting black holes}. pp 51--70

\bibitem[\protect\citeauthoryear{{Beniamini}, {Hotokezaka} \&
  {Piran}}{{Beniamini} et~al.}{2016}]{beniamini16a}
{Beniamini} P.,  {Hotokezaka} K.,    {Piran} T. 2016, ApJ, 832, 149

\bibitem[\protect\citeauthoryear{{Berger}, {Fong} \& {Chornock}}{{Berger}
  et~al.}{2013}]{berger13b}
{Berger} E.,  {Fong} W.,    {Chornock} R.  2013, ApJL, 774, L23

\bibitem[\protect\citeauthoryear{{Burbidge}, {Burbidge}, {Fowler} \&
  {Hoyle}}{{Burbidge} et~al.}{1957}]{burbidge57}
{Burbidge} E.~M.,  {Burbidge} G.~R.,  {Fowler} W.~A.,    {Hoyle} F.  1957,
  Reviews of Modern Physics, 29, 547

\bibitem[\protect\citeauthoryear{Cameron}{Cameron}{1957}]{cameron57}
Cameron A. G.~W. 1957, Chalk River Rept., CRL-41

\bibitem[\protect\citeauthoryear{{Ciolfi} \& {Siegel}}{{Ciolfi} \&
  {Siegel}}{2015}]{ciolfi15}
{Ciolfi} R.,  {Siegel} D.~M.  2015, ApJL, 798, L36
      
\bibitem[\protect\citeauthoryear{Coulter et al.}{2017}{}]{coulter17}
Coulter, D.A.~et al.  2017, Science 358, 1556 

\bibitem[Cowperthwaite et al.(2017)]{cowperthwaite17} Cowperthwaite,
P.~S., Berger, E., Villar, V.~A., et al.\ 2017, ApJ, 848, L17

\bibitem[\protect\citeauthoryear{{Dessart}, {Ott}, {Burrows}, {Rosswog} \&
  {Livne}}{{Dessart} et~al.}{2009}]{dessart09}
{Dessart} L.,  {Ott} C.~D.,  {Burrows} A.,  {Rosswog} S.,    {Livne} E.  2009,
  ApJ, 690, 1681

\bibitem[\protect\citeauthoryear{Duflo \& Zuker}{Duflo \&
  Zuker}{1995}]{duflo95}
Duflo J., Zuker A.~P.  1995, Phys. Rev. C, 52, R23

\bibitem[\protect\citeauthoryear{Eichler, Livio, Piran \& Schramm}{Eichler
  et~al.}{1989}]{eichler89}
Eichler D.,  Livio M.,  Piran T.,    Schramm D.~N.  1989, Nature, 340, 126

\bibitem[\protect\citeauthoryear{Evans et al. 2017}{}]{evans17}
Evans P. ~et al. 2017, Science 358, 1565 

\bibitem[\protect\citeauthoryear{{Fernandez} \& {Metzger}}{{Fernandez} \&
  {Metzger}}{2013}]{fernandez13a}
{Fernandez} R.,  {Metzger} B.~D.  2013, ApJ, 763, 108

\bibitem[\protect\citeauthoryear{{Fernandez}, {Kasen}, {Metzger} \& {Quataert}}
{{Fernandez} et al.}{2015}]{fernandez15}
{Fernandez} R., {Kasen} D.,  {Metzger} B.~D. \& {Quataert}, E. 2015,
MNRAS, 446, 750

\bibitem[\protect\citeauthoryear{Freiburghaus, Rembges, Rauscher, Kolbe,
  Thielemann, Kratz \& Cowan}{Freiburghaus et~al.}{1999}]{freiburghaus99a}
Freiburghaus C.,  Rembges J.,  Rauscher T.,  Kolbe E.,  Thielemann F.-K.,
  Kratz K.-L.,    Cowan J.  1999, ApJ, 516, 381

\bibitem[\protect\citeauthoryear{Freiburghaus, Rosswog \&
  Thielemann}{Freiburghaus et~al.}{1999}]{freiburghaus99b}
Freiburghaus C.,  Rosswog S.,    Thielemann F.-K.  1999, ApJ, 525, L121

\bibitem[\protect\citeauthoryear{{Goriely}, {Bauswein} \& {Janka}}{{Goriely}
  et~al.}{2011}]{goriely11a}
{Goriely} S.,  {Bauswein} A.,    {Janka} H.-T.  2011, ApJL, 738, L32

\bibitem[\protect\citeauthoryear{{Grossman}, {Korobkin}, {Rosswog} \&
  {Piran}}{{Grossman} et~al.}{2014}]{grossman14a}
{Grossman} D.,  {Korobkin} O.,  {Rosswog} S.,    {Piran} T.  2014, MNRAS, 439,
  757
 
\bibitem[Guillochon et al.(2017)]{guillochon17} Guillochon, J.,
Parrent, J., Kelley, L.~Z., \& Margutti, R.\ 2017, ApJ, 835, 64
  
\bibitem[\protect\citeauthoryear{{Hansen} et~al.}{2017}]{hansen17a}
{Hansen} T.T.,  et al.  2017, ApJ, 838, 44

\bibitem[\protect\citeauthoryear{{Hotokezaka}, Kiuchi, Kyutoku, Okawa, Sekiguchi \& {Shibata}}{{Hotokezaka}
  et~al.}{2013}]{hotokezaka13}
{Hotokezaka}, Kiuchi, Kyutoku, Okawa, Sekiguchi \& {Shibata}  2013, Phys. Rev. D, 87, 2

\bibitem[\protect\citeauthoryear{{Hotokezaka}, {Piran} \& {Paul}}{{Hotokezaka}
  et~al.}{2015}]{hotokezaka15a}
{Hotokezaka} K.,  {Piran} T.,    {Paul} M.  2015, Nature Physics, 11, 1042

\bibitem[\protect\citeauthoryear{{Hotokezaka}, {Sari} \& {Piran}}{{Hotokezaka}
  et~al.}{2017}]{hotokezaka17a}
{Hotokezaka} K.,  {Sari} R.,    {Piran} T.  2017, MNRAS, 468, 91

\bibitem[\protect\citeauthoryear{{Hulse} \& {Taylor}}{{Hulse} \&
  {Taylor}}{1975}]{hulse75}
{Hulse} R.~A.,  {Taylor} J.~H.  1975, ApJL, 195, L51

\bibitem[\protect\citeauthoryear{{Just}, {Bauswein}, {Pulpillo}, {Goriely} \&
  {Janka}}{{Just} et~al.}{2015}]{just15}
{Just} O.,  {Bauswein} A.,  {Pulpillo} R.~A.,  {Goriely} S.,    {Janka} H.-T.
  2015, MNRAS, 448, 541

\bibitem[\protect\citeauthoryear{{Kasen}, {Badnell} \& {Barnes}}{{Kasen}
  et~al.}{2013}]{kasen13a}
{Kasen} D.,  {Badnell} N.~R.,    {Barnes} J.  2013, ApJ, 774, 25

\bibitem[\protect\citeauthoryear{{Kasen}, {Fernandez} \& {Metzger}}{{Kasen}
  et~al.}{2015}]{kasen15a}
{Kasen} D.,  {Fernandez} R.,    {Metzger} B.D.  2015, MNRAS, 450, 1777

\bibitem[\protect\citeauthoryear{{Kasen}, {Metzger} \& {Barnes}}{{Kasen}
  et~al.}{2017}]{kasen17a}
{Kasen} D.,  {Metzger} B.D.,    {Barnes} J.,  Quataert, E., Ramirez-Ruiz, E. 2017, Nature, 551, 80

\bibitem[\protect\citeauthoryear{}{Kasliwal et al.}{2017}]{kasliwal17}
Kasliwal M.~et al. 2017, Science 358, 1559 

\bibitem[\protect\citeauthoryear{{Lai} et~al.}{{Lai}
  et~al.}{2008}]{lai08}
  {Lai}, D.~K. and {Bolte}, M. and {Johnson}, J.~A. and {Lucatello}, S. and 
{Heger}, A. and {Woosley}, S.~E.  2008, ApJ 681, 1524

\bibitem[\protect\citeauthoryear{}{Levan et al.}{2017}]{levan17}
Levan, A. et al. 2017, ApJ 848, L28

\bibitem[\protect\citeauthoryear{}{Lyman et al.}{2018}]{lyman18}
Lyman, J.D. et al., 2018, arXiv:1801.02669

\bibitem[\protect\citeauthoryear{}{Tanvir et al.}{2017}]{tanvir17}
N. Tanvir ~et al. 2017,  ApJ 848, L27 

\bibitem[\protect\citeauthoryear{}{Chornock et al.} {2017}]{chornock17}
Chornock, R.~et al. 2017, ApJ 848, L19

\bibitem[\protect\citeauthoryear{{Kim}, {Perera} \& {McLaughlin}}{{Kim}
  et~al.}{2015}]{kim15}
{Kim} C.,  {Perera} B.~B.~P.,    {McLaughlin} M.~A. 2015, MNRAS, 448, 928

\bibitem[\protect\citeauthoryear{{Korobkin}, {Rosswog}, {Arcones} \&
  {Winteler}}{{Korobkin} et~al.}{2012}]{korobkin12a}
{Korobkin} O.,  {Rosswog} S.,  {Arcones} A.,    {Winteler} C.  2012, MNRAS,
  426, 1940

\bibitem[\protect\citeauthoryear{{Kulkarni}}{{Kulkarni}}{2005}]{kulkarni05}
{Kulkarni} S.~R. 2005, arXiv:astro-ph/0510256 

\bibitem[\protect\citeauthoryear{Lattimer \& Schramm}{Lattimer \&
  Schramm}{1974}]{lattimer74}
Lattimer J.~M.,  Schramm D.~N.  1974, ApJ, 192, L145

\bibitem[\protect\citeauthoryear{{Lee} \& {Ramirez-Ruiz}}{{Lee} \&
  {Ramirez-Ruiz}}{2007}]{lee07}
{Lee} W.~H.,  {Ramirez-Ruiz} E.  2007, New Journal of Physics, 9, 17

\bibitem[\protect\citeauthoryear{{Li} \& {Paczy{\'n}ski}}{{Li} \&
  {Paczy{\'n}ski}}{1998}]{li98}
{Li} L.-X.,  {Paczy{\'n}ski} B.  1998, ApJL, 507, L59

\bibitem[\protect\citeauthoryear{Abbott et al. 2017}{Abbott et al.}{2017a}]{LIGO17_GW_NSNS}
Abbott, P.B., et al. 2017, Phys. Rev. Lett. 119, 161101

\bibitem[\protect\citeauthoryear{Margutti et al.}{2017}]{margutti17}
R. Margutti et al. 2017, ApJ, 848, L20

\bibitem[\protect\citeauthoryear{Moeller}{Moeller}{1995}]{moeller95}
P. M\"oller, J. R. Nix, W. D. Myers, W. J. Swiatecki 1995, At. Data Nucl. Data Tables, 59, 185

\bibitem[\protect\citeauthoryear{{Martin}, {Perego}, {Arcones}, {Thielemann},
  {Korobkin} \& {Rosswog}}{{Martin} et~al.}{2015}]{martin15}
{Martin} D.,  {Perego} A.,  {Arcones} A.,  {Thielemann} F.-K.,  {Korobkin} O.,
    {Rosswog} S.  2015, ApJ, 813, 2

\bibitem[\protect\citeauthoryear{{Matteucci}, {Romano}, {Arcones}, {Korobkin}
  \& {Rosswog}}{{Matteucci} et~al.}{2014}]{matteucci14a}
{Matteucci} F.,  {Romano} D.,  {Arcones} A.,  {Korobkin} O.,    {Rosswog} S.
  2014, MNRAS

\bibitem[\protect\citeauthoryear{{Mennekens} \& {Vanbeveren}}{{Mennekens} \&
  {Vanbeveren}}{2014}]{mennekens14a}
{Mennekens} N.,  {Vanbeveren} D.  2014, A \& A, 564, A134

\bibitem[\protect\citeauthoryear{{Metzger}, {Martinez-Pinedo}, {Darbha},
  {Quataert}, {Arcones}, {Kasen}, {Thomas}, {Nugent}, {Panov} \&
  {Zinner}}{{Metzger} et~al.}{2010}]{metzger10b}
{Metzger} B.~D.,  {Martinez-Pinedo} G.,  {Darbha} S.,  {Quataert} E.,
  {Arcones} A.,  {Kasen} D.,  {Thomas} R.,  {Nugent} P.,  {Panov} I.~V.,
  {Zinner} N.~T.  2010, MNRAS, 406, 2650

\bibitem[\protect\citeauthoryear{{Metzger}, {Piro} \& {Quataert}}{{Metzger}
  et~al.}{2008}]{metzger08a}
{Metzger} B.~D.,  {Piro} A.~L.,    {Quataert} E.  2008, MNRAS, 390, 781

\bibitem[\protect\citeauthoryear{{Metzger} \& {Fernandez}}{{Metzger}
  et~al.}{2014}]{metzger14}
 {Metzger} B.~D., {Fernandez}, R. 2014, MNRAS 441, 3444
 
 \bibitem[\protect\citeauthoryear{{Moesta} et~al.}{{Moesta}
  et~al.}{2017}]{moesta17}
 {Moesta} P. et al. 2017, arXiv:1712.09370
     
\bibitem[\protect\citeauthoryear{Goldstein et al.}{2017}{}]{fermiGBMpaper}
Goldstein  A. et al. 2017, ApJL 848, L14 

\bibitem[\protect\citeauthoryear{Hallinan et al.}{2017}{}]{vlapaper17}
Hallinan, G.~et al. 2017, Science 358, 1579 

\bibitem[\protect\citeauthoryear{{Perego}, {Gafton}, {Cabez{\'o}n}, {Rosswog}
  \& {Liebend{\"o}rfer}}{{Perego} et~al.}{2014}]{perego14a}
{Perego} A.,  {Gafton} E.,  {Cabez{\'o}n} R.,  {Rosswog} S.,
  {Liebend{\"o}rfer} M.  2014, A \& A, 568, A11

\bibitem[\protect\citeauthoryear{{Perego}, {Rosswog}, {Cabez{\'o}n},
  {Korobkin}, {K{\"a}ppeli}, {Arcones} \& {Liebend{\"o}rfer}}{{Perego}
  et~al.}{2014}]{perego14b}
{Perego} A.,  {Rosswog} S.,  {Cabez{\'o}n} R.~M.,  {Korobkin} O.,
  {K{\"a}ppeli} R.,  {Arcones} A.,    {Liebend{\"o}rfer} M.  2014, MNRAS, 443,
  3134

\bibitem[\protect\citeauthoryear{{Petrillo}, {Dietz} \& {Cavaglia}}{{Petrillo}
  et~al.}{2013}]{petrillo13}
{Petrillo} C.,  {Dietz} A.,    {Cavaglia} M.  2013, ApJ, 767, 140

\bibitem[Pian et al.(2017)]{pian17} Pian, E., D'Avanzo, P., Benetti, S., et al.\ 2017, Nature, 551, 67 

\bibitem[\protect\citeauthoryear{{Pinto} \& {Eastman}}{{Pinto} \&
  {Eastman}}{2000}]{pinto00a}
{Pinto} P.A.,  {Eastman} R.G. 2000, ApJ, 530, 744

\bibitem[\protect\citeauthoryear{{Quian} \& {Woosley}}{{Qian} \&
  {Woosley}}{1996}]{qian96b}
{Qian} Y.Z.,  {Woosley} S. 1996, ApJ, 471, 331

\bibitem[\protect\citeauthoryear{{Radice}, {Galeazzi}, {Lippuner}, {Roberts},
  {Ott} \& {Rezzolla}}{{Radice} et~al.}{2016}]{radice16a}
{Radice} D.,  {Galeazzi} F.,  {Lippuner} J.,  {Roberts} L.~F.,  {Ott} C.~D.,
  {Rezzolla} L.  2016, MNRAS, 460, 3255

\bibitem[\protect\citeauthoryear{{Roberts}, {Kasen}, {Lee} \&
  {Ramirez-Ruiz}}{{Roberts} et~al.}{2011}]{roberts11}
{Roberts} L.~F.,  {Kasen} D.,  {Lee} W.~H.,    {Ramirez-Ruiz} E.  2011, ApJL,
  736, L21

\bibitem[\protect\citeauthoryear{{Rosswog}}{{Rosswog}}{2005}]{rosswog05a}
{Rosswog} S.  2005, ApJ, 634, 1202

\bibitem[\protect\citeauthoryear{{Rosswog}}{{Rosswog}}{2013}]{rosswog13b}
{Rosswog} S. 2013, Royal Society of London Philosophical Transactions Series A, 371, 20272

\bibitem[\protect\citeauthoryear{{Rosswog}}{{Rosswog}}{2014}]{rosswog14a}
{Rosswog} S., {Korobkin}, O., {Arcones}, A., {Thielemann}, F.-K. and 
	{Piran}, T.  2014, MNRAS, 439, 744

\bibitem[\protect\citeauthoryear{Rosswog, Feindt, Korobkin, Wu, Sollerman,
  Goobar \& Martinez-Pinedo}{Rosswog et~al.}{2017}]{rosswog17a}
Rosswog S.,  Feindt U.,  Korobkin O.,  Wu M.-R.,  Sollerman J.,  Goobar A.,
  Martinez-Pinedo G.  2017, Classical and Quantum Gravity, 34, 104001

\bibitem[\protect\citeauthoryear{Rosswog, Liebend\"orfer, Thielemann, Davies,
  Benz \& Piran}{Rosswog et~al.}{1999}]{rosswog99}
Rosswog S.,  Liebend\"orfer M.,  Thielemann F.-K.,  Davies M.,  Benz W.,
  Piran T.  1999, A \&\ A, 341, 499

\bibitem[\protect\citeauthoryear{{Rosswog}, {Thielemann}, {Davies}, {Benz} \&
  {Piran}}{{Rosswog} et~al.}{1998}]{rosswog98b}
{Rosswog} S. et al. 1998, in {Hillebrandt} W.,  {Muller} E.,  
  Proceedings of the 9th workshop on Nuclear Astrophysics. Eds. Wolfgang Hillebrandt and Ewald M\"uller, p.~103,
  {Coalescing Neutron Stars: a Solution to the R-Process Problem?},
arXiv:astro-ph/9804332

\bibitem[\protect\citeauthoryear{{Shen}, {Cooke}, {Ramirez-Ruiz}, {Madau},
  {Mayer} \& {Guedes}}{{Shen} et~al.}{2015}]{shen15a}
{Shen} S.,  {Cooke} R.~J.,  {Ramirez-Ruiz} E.,  {Madau} P.,  {Mayer} L.,
  {Guedes} J.  2015, ApJ, 807, 115

\bibitem[\protect\citeauthoryear{{Siegel} \& {Metzger}}{{Siegel} \&
  {Metzger}}{2017}]{siegel17a}
{Siegel} D.~M.,  {Metzger} B.~D. 2017, Phys. Rev. Lett., 119, 23, id.231102

\bibitem[\protect\citeauthoryear{Smartt et al.}{2017}{}]{smartt17}
Smartt S.~J. e.~a. 2017, Nature 551, 75

\bibitem[\protect\citeauthoryear{{Sneden}, {Cowan} \& {Gallino}}{{Sneden}
  et~al.}{2008}]{sneden08}
{Sneden} C.,  {Cowan} J.~J.,    {Gallino} R.  2008, Annual Review of Astronomy
  and Astrophysics, 46, 241

\bibitem[\protect\citeauthoryear{{Tanaka} \& {Hotokezaka}}{{Tanaka} \&
  {Hotokezaka}}{2013}]{tanaka13a}
{Tanaka} M.,  {Hotokezaka} K.  2013, ApJ, 775, 113

\bibitem[\protect\citeauthoryear{{Tanvir}, {Levan}, {Fruchter}, {Hjorth},
  {Hounsell}, {Wiersema} \& {Tunnicliffe}}{{Tanvir} et~al.}{2013}]{tanvir13a}
{Tanvir} N.~R.,  {Levan} A.~J.,  {Fruchter} A.~S.,  {Hjorth} J.,  {Hounsell}
  R.~A.,  {Wiersema} K.,    {Tunnicliffe} R.~L.  2013, Nature, 500, 547

\bibitem[\protect\citeauthoryear{{Taylor} \& {Weisberg}}{{Taylor} \&
  {Weisberg}}{1982}]{taylor82}
{Taylor} J.~H.,  {Weisberg} J.~M.  1982, ApJ, 253, 908

\bibitem[\protect\citeauthoryear{{Thielemann}, {Arcones}, {K{\"a}ppeli},
  {Liebend{\"o}rfer}, {Rauscher}, {Winteler}, {Fr{\"o}hlich}, {Dillmann},
  {Fischer}, {Martinez-Pinedo}, {Langanke}, {Farouqi}, {Kratz}, {Panov} \&
  {Korneev}}{{Thielemann} et~al.}{2011}]{thielemann11}
{Thielemann} F.-K.,  {Arcones} A.,  {K{\"a}ppeli} R.,  {Liebend{\"o}rfer} M.,
  {Rauscher} T.,  {Winteler} C.,  {Fr{\"o}hlich} C.,  {Dillmann} I.,  {Fischer}
  T.,  {Martinez-Pinedo} G.,  {Langanke} K.,  {Farouqi} K.,  {Kratz} K.-L.,
  {Panov} I.,    {Korneev} I.~K.  2011, Progress in Particle and Nuclear
  Physics, 66, 346
  
  \bibitem[\protect\citeauthoryear{{Thielemann} et~al.}{{Thielemann}
  et~al.}{2017}]{thielemann17}
{Thielemann}, F.-K. and {Eichler}, M. and {Panov}, I.~V. and 
	{Wehmeyer}, B. 2017, Annual Review of Nuclear and Particle Science 67, 253
	
\bibitem[\protect\citeauthoryear{{van de Voort}, {Quataert}, {Hopkins}, {Keres}
  \& {Faucher-Giguere}}{{van de Voort} et~al.}{2015}]{vandevoort14a}
{van de Voort} F.,  {Quataert} E.,  {Hopkins} P.~F.,  {Keres} D.,
  {Faucher-Giguere} C.-A. 2015, MNRAS, 447, 140
  
\bibitem[Villar et al.(2017)]{villar17} Villar, V.~A., Guillochon, J., Berger, E., et al.\ 2017, ApJ, 851, L21

\bibitem[\protect\citeauthoryear{{Wallner}, {Faestermann}, {Feige},
  {Feldstein}, {Knie}, {Korschinek}, {Kutschera}, {Ofan}, {Paul}, {Quinto},
  {Rugel} \& {Steier}}{{Wallner} et~al.}{2015}]{wallner15a}
{Wallner} A.,  {Faestermann} T.,  {Feige} J.,  {Feldstein} C.,  {Knie} K.,
  {Korschinek} G.,  {Kutschera} W.,  {Ofan} A.,  {Paul} M.,  {Quinto} F.,
  {Rugel} G.,    {Steier} P.  2015, Nature Communications, 6, 5956

\bibitem[\protect\citeauthoryear{{Wanajo} \& {Janka}}{{Wanajo} \&
  {Janka}}{2012}]{wanajo12}
{Wanajo} S.,  {Janka} H.-T.  2012, ApJ, 746, 180

\bibitem[\protect\citeauthoryear{{Wanajo}, {Sekiguchi}, {Nishimura}, {Kiuchi},
  {Kyutoku} \& {Shibata}}{{Wanajo} et~al.}{2014}]{wanajo14}
{Wanajo} S.,  {Sekiguchi} Y.,  {Nishimura} N.,  {Kiuchi} K.,  {Kyutoku} K.,
  {Shibata} M. 2014, ApJL, 789, L39

\bibitem[\protect\citeauthoryear{{Winteler}}{{Winteler}}{2012}]{winteler12}
{Winteler} C. 2012, PhD thesis, University Basel, CH

\bibitem[\protect\citeauthoryear{{Winteler}, {K{\"a}ppeli}, {Perego},
  {Arcones}, {Vasset}, {Nishimura}, {Liebend{\"o}rfer} \&
  {Thielemann}}{{Winteler} et~al.}{2012}]{winteler12b}
{Winteler} C.,  {K{\"a}ppeli} R.,  {Perego} A.,  {Arcones} A.,  {Vasset} N.,
  {Nishimura} N.,  {Liebend{\"o}rfer} M.,    {Thielemann} F.-K. 2012, ApJL,
  750, L22

\bibitem[\protect\citeauthoryear{{Wollaeger}, {Korobkin}, {Fontes}, {Rosswog},
  {Even}, {Fryer}, {Sollerman}, {Hungerford}, {van Rossum} \&
  {Wollaber}}{{Wollaeger} et~al.}{2017}]{wollaeger17a}
{Wollaeger} R.~T.,  {Korobkin} O.,  {Fontes} C.~J.,  {Rosswog} S.~K.,  {Even}
  W.~P.,  {Fryer} C.~L.,  {Sollerman} J.,  {Hungerford} A.~L.,  {van Rossum}
  D.~R.,    {Wollaber} A.~B.  2017, arXiv:1705.07084
  
\bibitem[\protect\citeauthoryear{{Wu}, {Fernandez}, {Martinez-Pinedo} \&
  {Metzger}}{{Wu} et~al.}{2016}]{wu16}
{Wu} M.-R.,  {Fernandez} R.,  {Martinez-Pinedo} G., {Metzger} B.~D.  2016,
  MNRAS
    
\end{thebibliography}
\end{document}